\RequirePackage{silence}
\documentclass[fleqn,usenatbib,useAMS]{mnras} 
\usepackage{ragged2e}
\usepackage{graphicx}
\usepackage{threeparttable}
\usepackage{scrextend}
\usepackage{tablefootnote}
\usepackage{amsmath}

\usepackage{amsfonts}
\usepackage{float}
\usepackage{bm}
\usepackage{ae,aecompl}
\usepackage{array}
\usepackage{soul}
\usepackage{mathtools}
\usepackage{multirow}
\usepackage{bigdelim}
\newcolumntype{_}{>{\global\let\currentrowstyle\relax}}
\newcolumntype{^}{>{\currentrowstyle}}

\newcommand{\appropto}{\mathrel{\vcenter{
			\offinterlineskip\halign{\hfil$##$\cr 
				\propto\cr\noalign{\kern2pt}\sim\cr\noalign{\kern-2pt}}}}}
\defcitealias{Madau_Dickinson_2014}{MD14}

\title[Cosmological star formation history of the LV]{The cosmological star formation history from the Local Cosmological Volume of galaxies and constraints on the matter homogeneity} 

\author[M. Haslbauer et al.]{\parbox[t]{\textwidth} {Moritz Haslbauer$^{1}$\thanks{Email:
\href{mailto:mhaslbauer@astro.uni-bonn.de}{mhaslbauer@astro.uni-bonn.de} (Moritz Haslbauer)}, Pavel Kroupa$^{1,2}$ \& Tereza Jerabkova$^{3}$} \\
$^{1}$Helmholtz-Institut f\"ur Strahlen- und Kernphysik (HISKP), University of Bonn, Nussallee 14$-$16, D-53115 Bonn, Germany \\
$^{2}$Astronomical Institute, Faculty of Mathematics and Physics, Charles University, V Hole\v{s}ovi\v{c}k\'ach 2, CZ-180 00 Praha 8, Czech Republic\\
$^{3}$European Southern Observatory, Karl-Schwarzschild-Straße 2, D-85748 Garching bei M\"unchen, Germany \\}

\pubyear{2022}
\begin{document}
\label{firstpage}
\pagerange{\pageref{firstpage}--\pageref{lastpage}}

\maketitle

\begin{abstract}
The Lilly-Madau plot is commonly interpreted as the history of the cosmic star formation of the Universe by showing the co-moving star formation rate density (SFRD) over cosmic time. Therefore, the Lilly-Madau plot is sensitive not only to the star formation history (SFH) but also to the number density of galaxies. Assessing the Catalogue of Neighbouring Galaxies, we reconstruct the SFHs of galaxies located in the Local Cosmological Volume (LV) based on delayed-$\tau$ and power-law SFH models. Galaxies with stellar masses of $M_{*} \ga 10^{10}\,\rm{M_{\odot}}$ typically evolve according to the delayed-$\tau$ model by having first increasing followed by exponentially declining SFRs, while the majority of less massive star-forming galaxies have an almost constant or increasing SFH. Deducing the cosmic SFRD evolution of the LV reveals that the SFHs of local galaxies are inconsistent with the Lilly-Madau plot. The SFRDs of the LV are significantly lower at redshifts of $z \la 3$ underestimating the Lilly-Madau peak at $z = 1.86$ by a factor of $2.16\pm0.32$ (delayed-$\tau$ model) and $5.90\pm0.88$ (power-law model). Assuming the delayed-$\tau$ model for galaxies with $M_{*} \geq 10^{10}\,\rm{M_{\odot}}$ and a power-law model for less massive galaxies, the SFRD is $2.22\pm0.33$ lower than measured at $z = 1.86$. This inconsistency between the evolution of the local and global SFRD has cosmological implications.
Since the Lilly-Madau plot also constrains the cosmological matter field, the near-constancy of SFHs of LV galaxies could imply that the peak of the Lilly-Madau plot at $z = 1.86$ is the imprint of an $\approx~5$ co-moving Gpc-scale inhomogeneity.
\end{abstract}

\begin{keywords} 
galaxies: abundances -- galaxies: evolution -- galaxies: formation -- galaxies: star formation -- galaxies: stellar content -- large-scale structure of Universe
\end{keywords}

\section{Introduction} \label{sec:Introduction}
The observed history of cosmic star formation of the Universe constrains cosmological models of galaxy evolution and structure formation. Determining the stellar masses and star formation rates (SFRs) of galaxies over cosmic time is therewith of particular importance as the stellar mass growth reflects also the matter cycle and mass budget of the Universe \citep[e.g.][]{Kroupa_2020}.

Over the last three decades, several studies measured the SFR per co-moving volume of the Universe over cosmic time \citep[e.g.][]{Lilly_1996,Madau_1996,Madau_1998, Cole_2001, Madau_Dickinson_2014, Madau_2017} from which the following picture qualitatively emerged: The first stars and galaxies form during the epoch of the `cosmic dawn' at redshifts of $z \ga 9$ \citep{Laporte_2021}. This is followed by a phase in which galaxies rapidly grow in stellar mass such that the co-moving star formation rate density (SFRD) of the Universe increases from redshift $z \approx 8$ up to $z \approx 2$ by about one order of magnitude. During the so-called `high cosmic noon' at $2 \la z \la 3$, the SFRD becomes maximal which is widely interpreted as an active phase of the Universe where galaxies show high star formation activities. After that, the `cosmic afternoon' occurs during which the cosmic SFRD declines to the present-time by again about one order of magnitude. This evolution of the SFRD is the so-called Madau or Lilly-Madau plot and is understood as the history of the cosmic star formation of the Universe.

In a review, \citet[][hereafter \citetalias{Madau_Dickinson_2014}]{Madau_Dickinson_2014} show the evolution of the cosmic SFRD over redshift by measuring the SFRs in the rest-frame far-ultraviolet (FUV) and infrared (IR) continuum (see their figures~8 and 9) and assuming SFR-luminosity conversion factors for the Salpeter initial mass function \citep[IMF;][]{Salpeter_1955}. Their best fit to the cosmic co-moving SFRD \citepalias[see eq.~15 of][]{Madau_Dickinson_2014} has the form
\begin{eqnarray}
\centering
    SFRD_{\mathrm{MD14}}(z) = \frac{0.015 \, (1+z)^{2.7}}{1 + [(1+z)/2.9]^{5.6}} \, \rm{M_{\odot} \, yr^{-1}\, cMpc^{-3}} \, ,
    \label{eq:SFRD_Madau}
\end{eqnarray}
which yields an SFRD of $\approx 0.015 \, \rm{M_{\odot} \, yr^{-1}\, cMpc^{-3}}$ at $z = 0$ and a global maximum of $\approx 0.13 \, \rm{M_{\odot} \, yr^{-1}\, cMpc^{-3}}$ at $z = 1.86$ corresponding to an age of the Universe of $3.5\,\rm{Gyr}$ (see their figure~9 and also our Section~\ref{subsec:SFRD of the LV}). Thus, the SFRD increases by a factor of $\approx 8.9$ from the present time up to $z = 1.86$. The star formation history (SFH) and stellar mass growth of galaxies are physically linked such that the cosmic stellar mass density (SMD) can be deduced from the cosmic SFRD or vice versa. 

The Lilly-Madau plot as parametrized by Eq.~\ref{eq:SFRD_Madau} states the SFR per co-moving volume over cosmic time and therefore not only reflects the history of cosmic star formation but is also sensitive to the number density of the galaxies over cosmic time. Consequently, the Lilly-Madau plot also sets constraints on the matter homogeneity of the Universe. Galaxies in the local Universe provide the cosmological boundary conditions that have to be fulfilled by any viable model of structure formation and galaxy evolution \citep[][]{Kroupa_2020,Kroupa_Jerabkova_2021}. The sample of galaxies located in the Local Cosmological Volume \citep{Karachentsev_2004, Karachentsev_2013}, which is defined as a sphere with a radius of 11 Mpc centred around the Milky Way (MW), should have reliable well-observed quantities that can be assessed to constrain the cosmic SFH and matter density of the local Universe. While massive star-forming galaxies with stellar masses of $M_{\mathrm{*}} \ga 10^{10}\,\rm{M_{\odot}}$ typically follow the main sequence of star-forming galaxies \citep{Speagle_2014}, less massive galaxies show evidence for an almost constant SFR \citep{McGaugh_2017,Schombert_2019,Kroupa_2020} meaning that their present-day SFR, $SFR_{0}$, is comparable to the present-day averaged SFR, $\overline{SFR} \approx M_{*}/(12\,\rm{Gyr})$. Galaxies evolving according to the main sequence of star-forming galaxies \citep{Speagle_2014} have $SFR_{0}~<~\overline{SFR}$ and are described by the so-called delayed-$\tau$ SFH model in which the SFRs first increase and subsequently decline to present time (see their figures~9 and 10).

Analysing the Catalogue of Neighbouring Galaxies \citep{Karachentsev_2004, Karachentsev_2013}, \citet{Kroupa_2020} reported that the sample of galaxies in the Local Cosmological Volume (LV) with $SFR_{0} \geq 10^{-3}\, \rm{M_{\odot}\,yr^{-1}}$ has a mean value of $\overline{SFR}/SFR_{\mathrm{0}} = 0.85$ (see their figures~3-5) implying that most of the local galaxies have nearly constant SFRs over cosmic time. The SFH of these galaxies can be parametrized by a power-law of the form $SFR(t) \propto (t-t_{\mathrm{start}})^{\eta}$ for $t_{\mathrm{start}} = 1.8\,\rm{Gyr}$ and $\eta = 0.18 \pm 0.03$. As discussed by \citet{Kroupa_2020}, such an SFH is inconsistent with the evolutionary track of galaxies that strictly evolve along the main sequence of star-forming galaxies by \citet[][]{Speagle_2014}. Galaxies on the main sequence typically have $\overline{SFR}/SFR_{\mathrm{0}} \approx 2$ \citep{Kroupa_2020}. Similar results have been found by \citet{Schombert_2019}, who showed that galaxies of the \citet{Cook_2014} and LSB + SPARC \citep[][]{McGaugh_2017} sample with $M_{*} \la 10^{10} \, \rm{M_{\odot}}$ are consistent with a constant SFH over a time-scale of 13~Gyr (see their figure~1).  

Given that the majority of galaxies located in the LV have almost constant SFRs over cosmic time \citep{Kroupa_2020}, the question arises whether the SFHs of local galaxies are consistent with the global history of cosmic star formation as implied by the Lilly-Madau plot. In this contribution, we supplement the work by \citet{Kroupa_2020} by investigating the SFRs and stellar masses of galaxies located in the LV \citep{Karachentsev_2004, Karachentsev_2013} in the context of the Lilly-Madau plot. The paper is organized as follows: Section~\ref{sec:Methods} describes the observational data extracted from the updated version of the Catalogue of Neighbouring Galaxies as compiled by \citet{Karachentsev_2004,Karachentsev_2013} and introduces the parametrization of different SFH models. The SFHs, SFRD, and SMD of the LV are presented and compared with the Lilly-Madau plot in Section~\ref{sec:Results}. A discussion on the SFRD and matter homogeneity followed by concluding remarks are provided, respectively, in Sections~\ref{sec:Discussion} and \ref{sec:Conclusion}. Throughout the analysis we assume a standard Lambda cold dark matter ($\Lambda$CDM) cosmology with a global Hubble constant of $H_{0} = 67.74\,\rm{km\,s^{-1}\,Mpc^{-1}}$, and a present-day total matter density and dark energy density in units of the cosmic critical density of $\Omega_{\mathrm{m},0} = 0.3089$ and $\Omega_{\Lambda,0} = 0.6911$, respectively \citep{Planck_2016_IllustrisTNG}.

\section{Methods}
\label{sec:Methods}
This section introduces the observational parameters of the sample of neighbouring galaxies and the SFH models used to calculate the cosmic SFRD and SMD of the LV.

\subsection{The Local Cosmological Volume} \label{subsec:Data}
The Local Cosmological Volume (hereafter LV) is defined as a sphere with a radius of $11\,\rm{Mpc}$ centred around the MW. The observational data of galaxies located inside the LV are extracted from the updated version \citep{Karachentsev_2013} of the Catalogue of Neighbouring Galaxies \citep{Karachentsev_2004}\footnote{The catalogues of galaxies in the LV can be downloaded using the following link: \url{https://www.sao.ru/lv/lvgdb/introduction.php}. Here, we use the latest update from 12.04.2023.}, which lists galaxies with a Galactic-centric distance of $D < 11 \,\rm{Mpc}$ or radial velocities of $V < 600 \, \rm{km\,s^{-1}}$ \citep[see section~2 of][]{Karachentsev_2013}. We extract the $Ks$-band luminosities, $L_{\mathrm{Ks}}$, the Galactocentric distances, $D$, and the SFRs based on the integrated H$\alpha$ and FUV measurements \citep{Karachentsev_2013_SFRproperties}, $SFR_{\mathrm{H\alpha}}$ and $SFR_{\mathrm{FUV}}$, respectively (including flagged SFRs). Selecting galaxies with $D < 11 \, \rm{Mpc}$ results in a total sample of $1102$ galaxies from which $115$ have only the $SFR_{\mathrm{H\alpha}}$, $447$ only the $SFR_{\mathrm{FUV}}$, and $540$ both SFR measurements.

\subsubsection{Present-day SFR} \label{subsubsec:Present-day SFR}
The present-day SFR of galaxies with both SFR measurements is calculated by
\begin{eqnarray}
    \centering
    SFR_{0}~=~\frac{SFR_{H\alpha} + SFR_{\mathrm{FUV}}}{2} \, ,
    \label{eq:SFR_present}
\end{eqnarray}
and $SFR_{0} = SFR_{H_{\alpha}}$ or $SFR_{0} = SFR_{\mathrm{FUV}}$ is adopted if only $H_{\alpha}$- or FUV-based measurements are available, respectively \citep[see also section~2 of][]{Kroupa_2020}. 

\subsubsection{Present-day SFRD and SMD} \label{subsubsec:Present-day SFRD}
The present-day SFRD of the LV is
\begin{eqnarray}
    \centering
    SFRD(D \leq 11 \, \mathrm{Mpc})~\equiv~\frac{1}{V} \sum_{i=1}^{N} SFR_{0,i} \, ,
    \label{eq:SFRD_LV_0}
\end{eqnarray}
where $N$ is the number of galaxies located in a co-moving volume $V$. The estimated lower limit of the SFRD is simply given for the volume of a sphere with a radius of $11 \, \rm{Mpc}$
\begin{eqnarray}
    \centering
    V~=~\frac{4 \pi (11\,\rm{Mpc)}^{3}}{3} \approx 5575.3\,\rm{Mpc^3} \, ,
    \label{eq:Volume_maximum}
\end{eqnarray}
while the upper limit corresponds to a volume of 
\begin{eqnarray}
    \centering
    V~=~\frac{4 \pi (11\,\rm{Mpc})^{3}}{3} \big(1 - \sin(15^{\circ}) \big) \approx 4132.3 \, \rm{Mpc^3}\, ,
    \label{eq:Volume_minimum}
\end{eqnarray}
by taking into account potentially missed galaxies due to dust obscuration at Galactic latitudes of $b < \left | 15^{\circ} \right |$ \citep[e.g. section~2 of][]{Karachentsev_2018}. This gives an SFRD of $0.014-0.019\, \rm{M_{\odot} \, yr^{-1} \, Mpc^{-3}}$ for the LV, which is consistent with the present-day SFRD of $0.015\, \rm{M_{\odot} \, yr^{-1} \, Mpc^{-3}}$ as found by \citetalias{Madau_Dickinson_2014}. 

The present-day SMD in the LV is
\begin{eqnarray}
    \centering
    SMD(D \leq 11 \, \mathrm{Mpc})~\equiv~\frac{1}{V} \sum_{i=1}^{N} M_{*,i} \, ,
    \label{eq:SMD_LV_0}
\end{eqnarray}
where $M_{*,i}$ is the present-day stellar mass of the $i$-th galaxy assuming a mass-to-light ratio of $0.6\,\rm{M_{\odot}/L_{\odot}}$ in the $Ks$-band \citep[e.g.][]{McGaugh_2014}. The LV has an SMD of $(3.45-4.65)\times10^{8}~\,~\rm{M_{\odot}\,Mpc^{-3}}$. 

\subsubsection{Present-day averaged SFR} \label{subsubsec:Present-day averaged SFR}
Throughout the analysis we assume that all galaxies start forming stars $0.2$~Gyr after the big bang\footnote{We note that this differs from the analysis by \citet[][]{Kroupa_2020} in which $t_{\mathrm{start}} = 1.8 \, \rm{Gyr}$.}, which corresponds to a redshift of $z=18$ in standard $\Lambda$CDM cosmology \citep{Planck_2016_IllustrisTNG}. The observed present-day averaged SFR over the onset of SFR at time $t_{\mathrm{start}} = 0.2$~Gyr and present-day $t_{\mathrm{end}} = 13.8 \,\rm{Gyr}$ is then defined by 
\begin{eqnarray}
    \centering
    \overline{SFR}~\equiv~\frac{\zeta M_{*}}{t_{\mathrm{end}} - t_{\mathrm{start}}} = ~\frac{\zeta M_{*}}{t_{\mathrm{sf}}} \, ,
    \label{eq:SFR_averaged}
\end{eqnarray}
where $\zeta$ takes into account the mass-loss through stellar evolution. Simulations conducted by \citet{Baumgardt_2003} suggested a stellar mass loss from bound stars of about $30\%$ implying $\zeta = 1.3$ for the canonical two-part power-law IMF \citep{Kroupa_2001}. Throughout this study, we also set $\zeta = 1.3$ in order to be consistent with the analysis of \citet{Kroupa_2020}.

\subsection{Delayed-$\tau$ SFH model} \label{subsec:delayedtau model}
The delayed-$\tau$ model describes the SFH of a galaxy assuming that the SFRs typically rise in the early phase of galaxy evolution and gradually decline to the present time \citep[e.g.][]{Reddy_2012, Carnall_2019}. In fact, \citet{Speagle_2014} showed in their figures~9 and 10 that the SFH of galaxies following the main sequence of star-forming galaxies can be accurately parametrized by the delayed-$\tau$ model of the form
\begin{eqnarray}
&&SFR_{\mathrm{del}}(t)= \\
&&=\!\left\{
\begin{array}{ll}\frac{A_{\mathrm{del}}}{\tau^{2}}(t-t_{\mathrm{start}})\exp\bigg(-\frac{t-t_{\mathrm{start}}}{\tau}\bigg) \! & \text{if} \, t > t_{\mathrm{start}} \, , \\
 \!0 & \text{if} \, t \leq t_{\mathrm{start}} \, , \nonumber \\
\end{array}
\right.
\label{eq:delayedtau}
\end{eqnarray}
where $t$ is the age of the Universe, $A_{\mathrm{del}}$ is the normalization constant, and $\tau$ is the star formation time-scale. The model yields a declining SFH for $t/\tau \gg 1$ and a linear rising SFH for $t/\tau \ll 1$ \citep[see also e.g. section~3.1.4 of ][]{Speagle_2014}. Empirically, galaxies on the main sequence have $3.5 \la \tau/\rm{Gyr} \la 4.5$ \citep{Speagle_2014,Kroupa_2020}. We also allow solutions with $\tau < 0$ in order to describe galaxies with a gradually increasing SFH. 

\begin{figure*}
    \includegraphics[width=\linewidth]{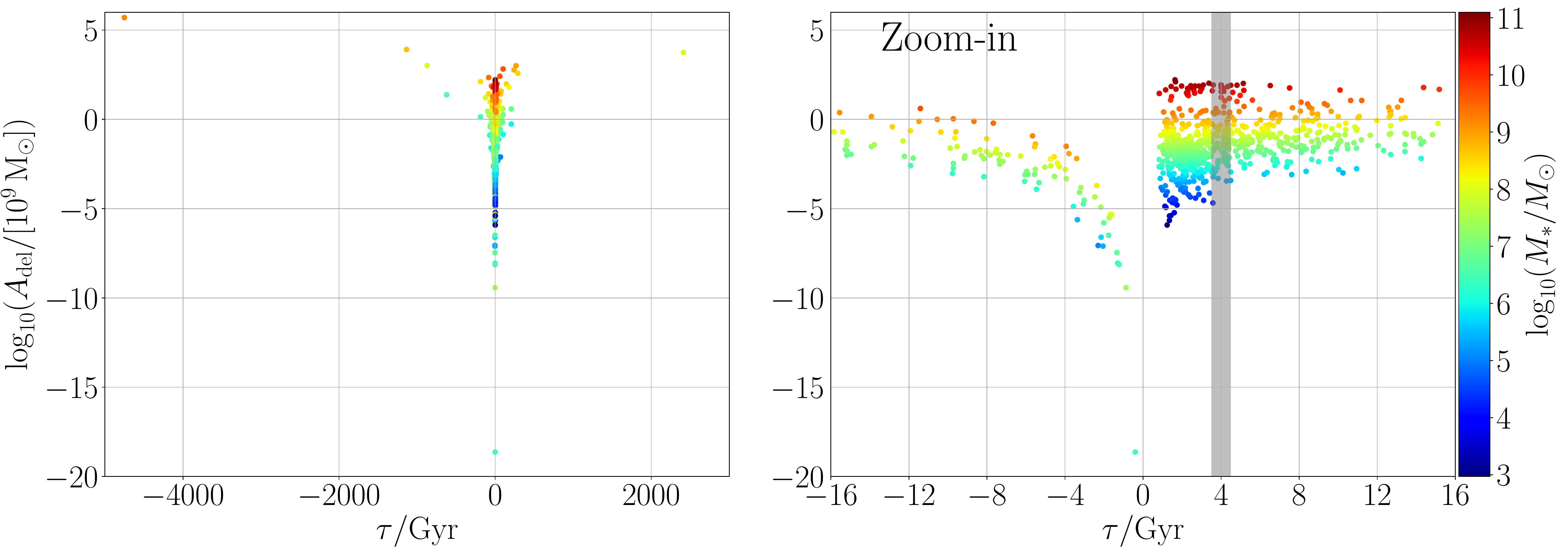}
    \caption{Parameter distribution of the delayed-$\tau$ SFH models for galaxies located in the LV. The colourbar represents the present-day stellar mass of the galaxies. The left-hand panel shows the parameter range of the LV galaxies (apart from one galaxy with an extreme negative value of $\tau$), while the right-hand panel is a zoom-in on galaxies with $-16 \leq \tau/\rm{Gyr} \leq 16$. Galaxies on the main sequence have $3.5 \la \tau/\rm{Gyr} \la 4.5$ \citep{Speagle_2014,Kroupa_2020} as highlighted by the grey area. Galaxies with $\tau < 0$ have slowly risings SFRs.} \label{figure_parameterspace_delayedTau_model}
\end{figure*}

The present-day SFR is given by adopting $t = t_{\mathrm{end}} = 13.8\,\rm{Gyr}$, such that $t_{\mathrm{sf}} = 13.6$~Gyr is the total star-forming lifetime,
\begin{eqnarray}
    SFR_{0,\mathrm{del}}~\equiv~SFR_{\mathrm{del}}(t=t_{\mathrm{end}}) = \frac{A_{\mathrm{del}} t_{\mathrm{sf}}}{\tau^{2}} \exp\bigg(-\frac{t_{\mathrm{sf}}}{\tau}\bigg) \, .    \label{eq:delayedtau_present}
\end{eqnarray}
The present-day averaged SFR is
\begin{eqnarray}
    \overline{SFR}_{\mathrm{del}}&\equiv&\frac{1}{t_{\mathrm{sf}}}\int_{t_{\mathrm{start}}}^{t_{\mathrm{end}}} SFR_{\mathrm{del}}(t) dt \nonumber \\ &=& \frac{A_{\mathrm{del}}}{t_{\mathrm{sf}}} \bigg[1-\bigg(1 + \frac{t_{\mathrm{sf}}}{\tau}\bigg) \exp\bigg(-\frac{t_{\mathrm{sf}}}{\tau}\bigg)\bigg] \, .    \label{eq:delayedtau_averaged}
\end{eqnarray}
Solving these two equations for the observed present-day (Section~\ref{subsubsec:Present-day SFR}) and present-day averaged (Section~\ref{subsubsec:Present-day averaged SFR}) SFRs of the LV galaxies, determines the parameters $A_{\mathrm{del}}$ and $\tau$. This allows to reconstruct the SFH of each LV galaxy as illustrated for some selected galaxies in Section~\ref{subsec:SFH of galaxies in the LV}. Figure~\ref{figure_parameterspace_delayedTau_model} shows the distribution of the parameters $A_{\mathrm{del}}$ and $\tau$ of all LV galaxies.  

Reconstructing the SFHs of individual LV galaxies enables to calculate the evolution of the SFRD over cosmic time. The SFRD of the LV based on the delayed-$\tau$ model is therefore
\begin{eqnarray}
    SFRD_{\mathrm{del}}(t) = \frac{1}{V}\, \sum_{i= 1}^{N} SFR_{\mathrm{del,i}}(t) \, , \label{eq:delayedtau_SFRD}
\end{eqnarray}
where $SFR_{\mathrm{del,i}}(t)$ is the SFR at time $t$ for the $i$-th galaxy. The so-calculated time evolution of the SFRD is presented in Section~\ref{subsec:SFRD of the LV}. 

\subsection{Power-law SFH model} \label{subsec:eta model} 
\begin{figure}
    \includegraphics[width=\linewidth]{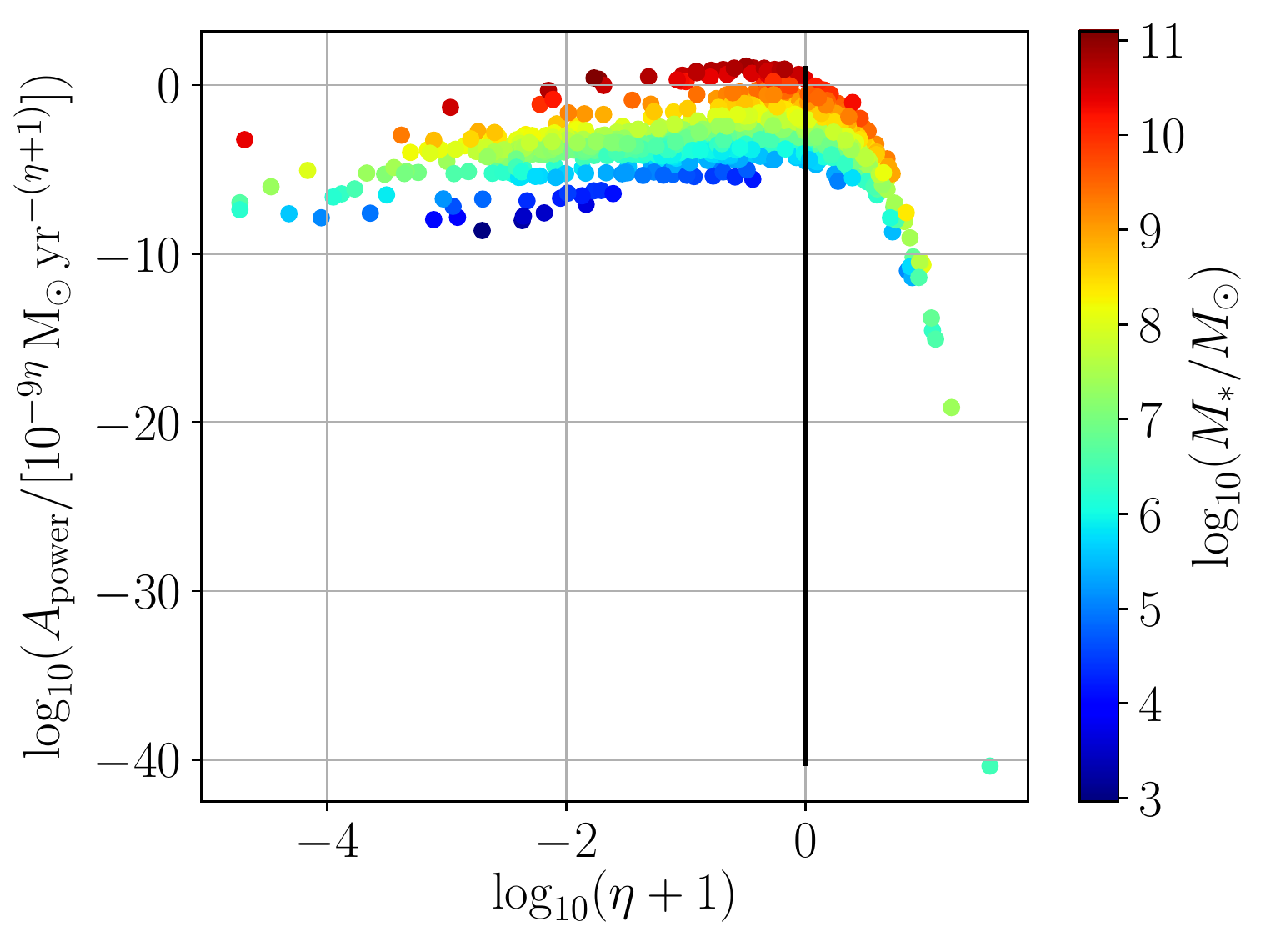}
    \caption{Parameter distribution of the power-law-$\eta$ SFH model for galaxies located in the LV. The colourbar represents the present-day stellar mass of the shown galaxies. The black solid line separates galaxies with an rising (i.e. $\eta > 0$) and a falling (i.e. $\eta < 0$) SFH.}
    \label{figure_parameterspace_eta_model}
\end{figure}

Following section~3.3 of \citet{Kroupa_2020}, the SFHs of LV galaxies can be parametrized by a power-law (hereafter power-law-$\eta$ SFH model) of the form

\begin{equation}
    \centering
    SFR_{\mathrm{power}}(t)~=~\left\{
\begin{array}{ll}
 A_{\mathrm{power}} (t-t_{\mathrm{start}})^{\eta} \, & \text{if} \quad t > t_{\mathrm{start}} \, , \\
 0\, \, & \text{if} \quad t \leq t_{\mathrm{start}} \, , \\
\end{array}
\right.
    \label{eq:etaSFHmodel}
\end{equation}
where $A_{\mathrm{power}}$ is the normalization constant and $\eta~\neq~1$ is the power-law index such that $\eta~=~0$, $>~0$, and $<~0$ implies a constant, increasing, and decreasing SFH, respectively. The present-day SFR is given by
\begin{eqnarray}
    SFR_{0,\mathrm{power}}~\equiv~SFR_{\mathrm{power}}(t = t_{\mathrm{end}})~=~A_{\mathrm{power}} (t_{\mathrm{sf}})^{\eta} \, ,
    \label{eq:eta_present}
\end{eqnarray}
and the present-day averaged SFR is
\begin{eqnarray}
    \overline{SFR}_{\mathrm{power}}~\equiv~\frac{1}{t_{\mathrm{sf}}}\int_{t_{\mathrm{start}}}^{t_{\mathrm{end}}} SFR_{\mathrm{power}}(t) dt~= ~\frac{A_{\mathrm{power}}}{1 + \eta} t_{\mathrm{sf}}^{\eta} \, .
    \label{eq:eta_averaged}
\end{eqnarray}
It follows that the power-law index depends on the present-day and present-day averaged SFR via
\begin{eqnarray}
    \eta = \frac{SFR_{0,\mathrm{power}}}{\overline{SFR}_{\mathrm{power}}} - 1 \, .    \label{eq:eta_parameter}
\end{eqnarray}
Solving the Equations~\ref{eq:eta_present} and \ref{eq:eta_averaged} for the present-day and present-day averaged SFRs of the LV sample, yields their parametrization of the power-law-$\eta$ SFH model as defined by Eq.~\ref{eq:etaSFHmodel}. The distribution of the parameters $A_{\mathrm{power}}$ and $\eta$ of the LV galaxies is depicted in Figure~\ref{figure_parameterspace_eta_model}. 

As elaborated in the previous section, the cosmic SFRD of the LV based on the power-law-$\eta$ SFH model is
\begin{eqnarray}
    SFRD_{\mathrm{power}}(t) = \frac{1}{V}\, \sum_{i=1}^{N} SFR_{\mathrm{power,i}}(t) \, .    \label{eq:eta_SFRD}
\end{eqnarray}

\subsection{Combined SFH model} \label{subsec:Combined SFH model}
Star-forming galaxies with $M_{*} \ga 10^{10} \, \rm{M_{\odot}}$ typically follow the main sequence of \citet{Speagle_2014} while most galaxies with $M_{*} \la 10^{10} \, \rm{M_{\odot}}$ have an almost constant SFH \citep[][]{McGaugh_2017,Schombert_2019, Kroupa_2020}. Therefore, we also apply a combination of the delayed-$\tau$ and power-law-$\eta$ SFH model by adopting, 
\begin{eqnarray}
    SFR_{\mathrm{combined}}(t)~=~\left\{
\begin{array}{ll} SFR_{\mathrm{del}}(t) \, & \text{if} \, M_{*} \geq 10^{10}\,\rm{M_{\odot}}  \, , \\
 SFR_{\mathrm{power}}(t) \, & \text{if} \, M_{*} < 10^{10}\,\rm{M_{\odot}} \, , \\
\end{array}
\right.
\label{eq:SFHscombined}
\end{eqnarray}
and
\begin{eqnarray}    SFRD_{\mathrm{combined}}(t) = \frac{1}{V}\, \sum_{i=1}^{N} SFR_{\mathrm{combined,i}}(t) \, .    \label{eq:SFRD_combined}
\end{eqnarray}
Throughout the study we refer to this as the combined SFH model. 

\subsection{Stellar mass density} \label{subsec:SMD} 
The cosmic evolution of the co-moving SFRD and SMD are physically connected via
\begin{eqnarray}
SMD(t) =  (1-R) \int_{0}^{t} SFRD(t') dt'\, ,  \label{eq:SMD_time}
\end{eqnarray}
which rewrites in terms of redshifts to
\begin{eqnarray}
SMD(z) =  (1-R) \int_{z}^{\infty} SFRD(z') \frac{dz'}{H(z') (1+z')}\, ,    \label{eq:SMD_redshift}
\end{eqnarray}
where $H(z) = H_{0} \sqrt{\Omega_{\mathrm{m,0}} (1 + z)^{3}+ (1 - \Omega_{\mathrm{m,0}})}$ is the Hubble parameter in a standard $\Lambda$CDM cosmology \citep{Planck_2016_IllustrisTNG} and $R = 1 - 1/\zeta$ \citep[e.g. section~4.1 of][]{Pipino_2014} is the recycling fraction \citep[see also section~4 of][]{Yu_2016}.

\section{Results}
\label{sec:Results}
\begin{figure*}
    \includegraphics[width=\linewidth]{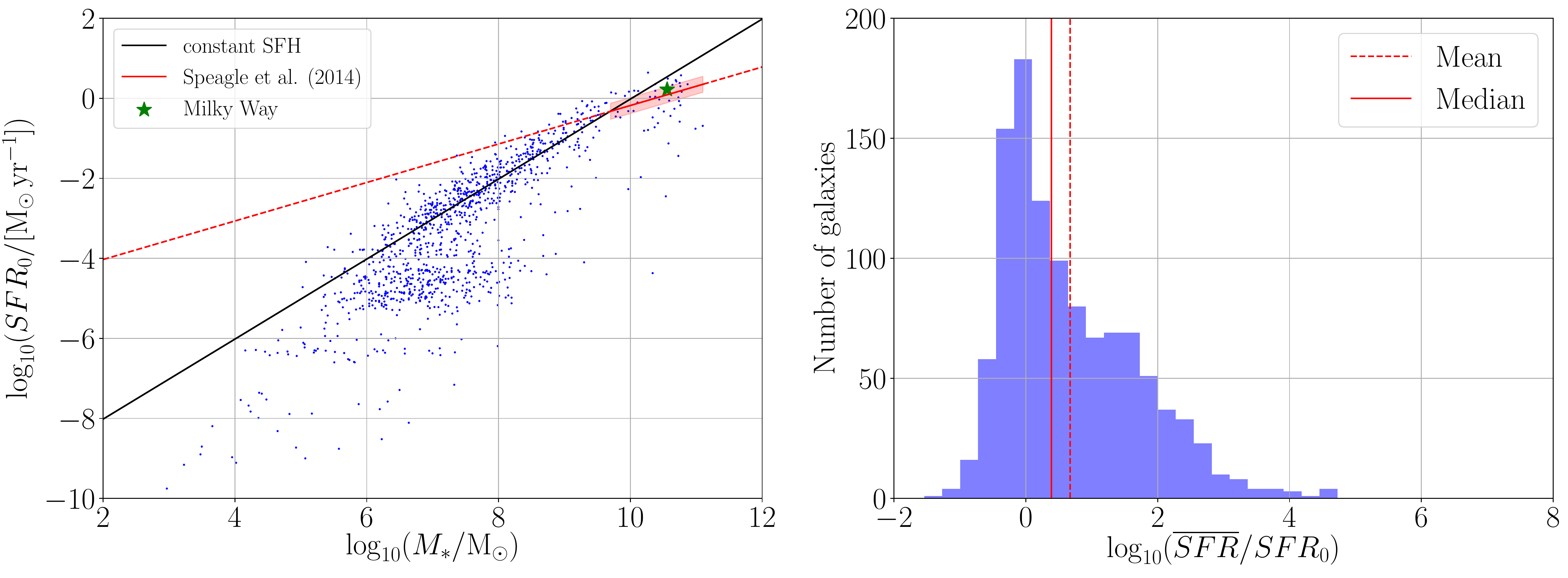}
    \caption{Left: Present-day SFR, $SFR_{0}$, in dependence of the stellar mass of galaxies located in the LV. The black line marks a constant SFH by assuming $\zeta = 1.3$ and $t_{\mathrm{sf}} = 13.6$~Gyr (Equation~\ref{eq:SFR_averaged}). The red solid line shows the present-day main sequence of star-forming galaxies with $10^{9.7}\leq M_{*}/M_{\odot}\leq 10^{11.1}$ \citep[eq.~28 and fig.~8 of ][]{Speagle_2014} and an uncertainty of $\pm0.2$~dex (red shaded region). The red dashed line extrapolates the main sequence beyond the adopted fitting range by \citet{Speagle_2014}. The green star marks the MW with $M_{*} = 3.6\times10^{10}\,\rm{M_{\odot}}$ and $SFR_{0} = 1.7\,\rm{M_{\odot}\,yr^{-1}}$ \citep{Karachentsev_2013}. Right: Distribution of $\overline{SFR}/SFR_{0}$ values of galaxies in the LV with a median and mean value of $0.39$ (solid line) and $0.67$ (dashed line) in the $\log_{10}$-space, respectively.}
    \label{figure_Mainseqeuence}
\end{figure*}

The present-day SFR (Section~\ref{subsubsec:Present-day SFR}) in dependence of the stellar mass of galaxies located in the LV is presented in the left-hand panel of Figure~\ref{figure_Mainseqeuence}. Low-mass galaxies with $M_{*}~\la~10^{10} \,\rm{M_{\odot}}$ and $SFR_{0} \ga  10^{-3}\,\rm{M_{\odot}\,yr^{-1}}$ typically have a constant SFH (black line), while galaxies with $M_{*}\ga 10^{10} \,\rm{M_{\odot}}$ have $SFR_{0} < \overline{SFR}$ following the main sequence of star-forming galaxies (solid red line). The right-hand panel shows the distribution of the values $\overline{SFR}/SFR_{0}$ in the log$_{10}$-space, which peaks at $\overline{SFR}/SFR_{0}\approx 1$ and has mean (median) of $4.69$ ($2.45$). Galaxies on the main sequence are expected to have $\overline{SFR}/SFR_{0} \approx 2$ \citep{Speagle_2014,Kroupa_2020}.

For a comparison with \citet{Kroupa_2020}, we select only galaxies with $SFR_{0} \geq 10^{-3}\,\rm{M_{\odot}\,yr^{-1}}$ which gives a sample size of $511$. These galaxies have a mean (median) value of $\overline{SFR}/SFR_{0} = 0.97$ ($0.83$) being broadly consistent with \citet{Kroupa_2020}\footnote{We note that \citet{Kroupa_2020} used the Catalogue of Neighbouring Galaxies from 08.05.2020.} who obtained a mean (median) of $1.11$ ($0.96$) for $t_{\mathrm{end}} = 1.8$~Gyr. 

 \subsection{SFHs of galaxies in the LV} \label{subsec:SFH of galaxies in the LV}
\begin{figure*}
    \includegraphics[width=\linewidth]{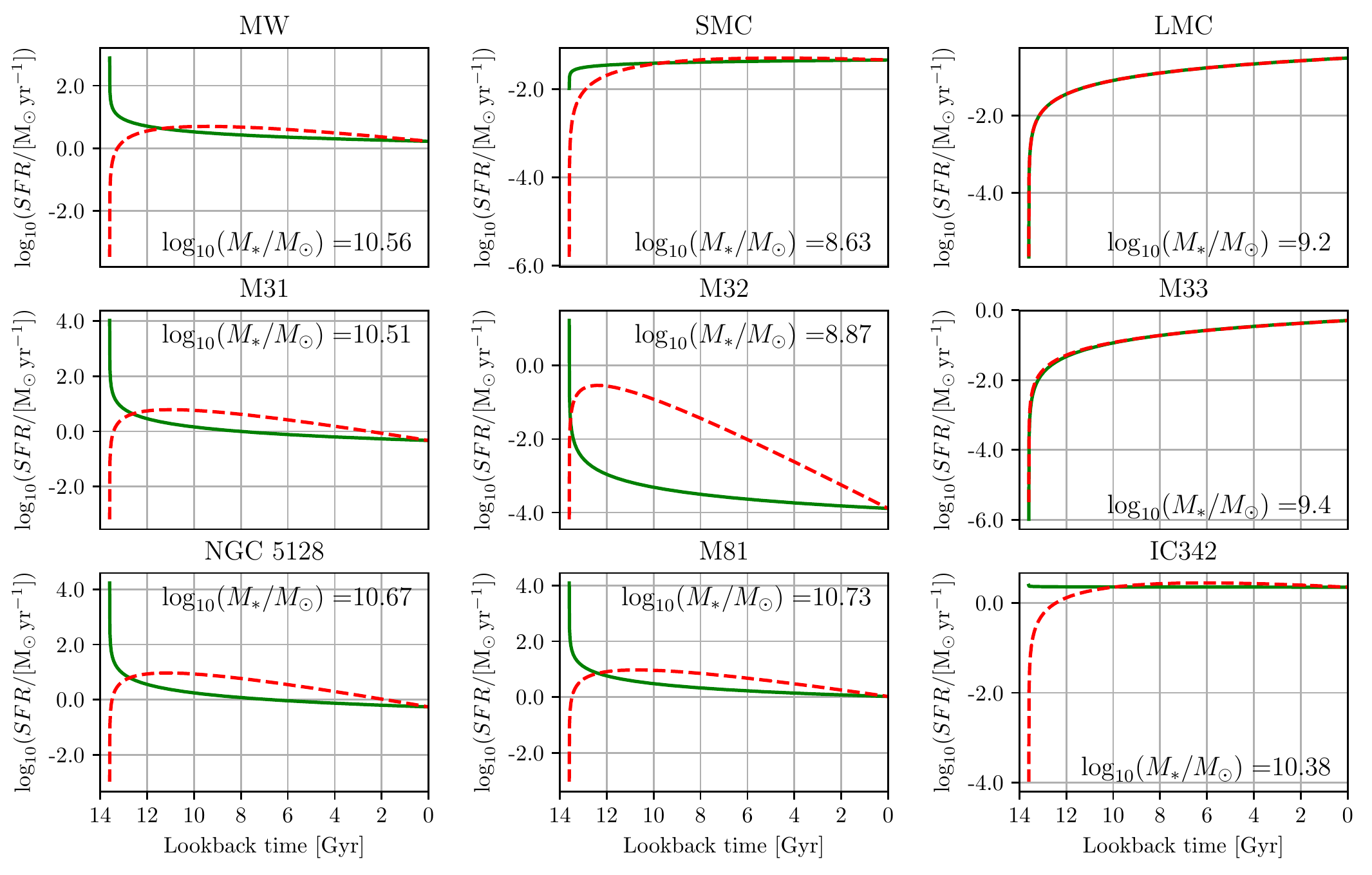}
    \caption{Reconstructed SFHs of the LV galaxies MW, SMC, LMC, M31, M32, M33, NGC 5128/Cen A, M81, and IC 342 assuming the delayed-$\tau$ (dashed red) and power-law-$\eta$ (solid green) SFH model. The panels state the present-day stellar masses of the galaxies. Note that the delayed-$\tau$ and power-law-$\eta$ SFH model give basically the same SFH for the LMC (top right) and M33 (middle right) galaxy.}
    \label{figure_SFHs}
\end{figure*}

The SFHs of selected galaxies in the LV parametrized by the delayed-$\tau$ (Section~\ref{subsec:delayedtau model}) and the power-law-$\eta$ (Section~\ref{subsec:eta model}) SFH models are presented in Figure~\ref{figure_SFHs}. In the delayed-$\tau$ model, the massive galaxies (i.e. $M_{*} > 10^{10}\,\rm{M_{\odot}}$) MW, M31, NGC~5128/Cen~A, M81, and IC~342 have first a phase of increasing SFRs, which is followed by a decreasing SFH over cosmic time. For example, the SFR of the MW globally peaks at a lookback time of $\approx9.5$~Gyr with a value of $5 \,\rm{M_{\odot}\,yr^{-1}}$ and subsequently decreases to a present-day value of $1.7 \,\rm{M_{\odot}\,yr^{-1}}$. The less massive Small Magellanic Cloud (SMC) and M32 galaxy with a stellar mass of $4.2\times10^8\,\rm{M_{\odot}}$ and $7.4\times10^8\,\rm{M_{\odot}}$, respectively, have also first an increasing and later on a decreasing SFH. The SMC has a maximal SFR of $0.05\,\rm{M_{\odot}\,yr^{-1}}$ at a lookback time of $\approx 4.7~\rm{Gyr}$, which is about $1.1$ times higher than the present-day value. M32 has an SFR peak of $0.3 \,\rm{M_{\odot}\,yr^{-1}}$ at a lookback time of $12.4$~Gyr and a present-day SFR of only $1.3 \times 10^{-4}\,\rm{M_{\odot}\,yr^{-1}}$. The Large Magellanic Cloud (LMC) with $M_{*} = 1.6\times 10^{9}\,\rm{M_{\odot}}$ and M33 with $M_{*} = 2.5\times 10^{9}\,\rm{M_{\odot}}$ have continuously rising SFHs in both the delayed-$\tau$ and power-law-$\eta$ SFH model. 

In the power-law-$\eta$ SFH model, the massive galaxies MW, M31, NGC~5128/Cen~A, M81, IC~342, but also M32 have a decreasing, while the low-mass galaxies SMC, LMC, and M33 have an increasing SFH over cosmic time in agreement with e.g. \citet[][]{McGaugh_2017}, \citet{Schombert_2019}, and \citet{Kroupa_2020}. 

\subsection{Cosmic SFRD and SMD of the LV} \label{subsec:SFRD of the LV}
\begin{figure*}
    \includegraphics[width=\linewidth]{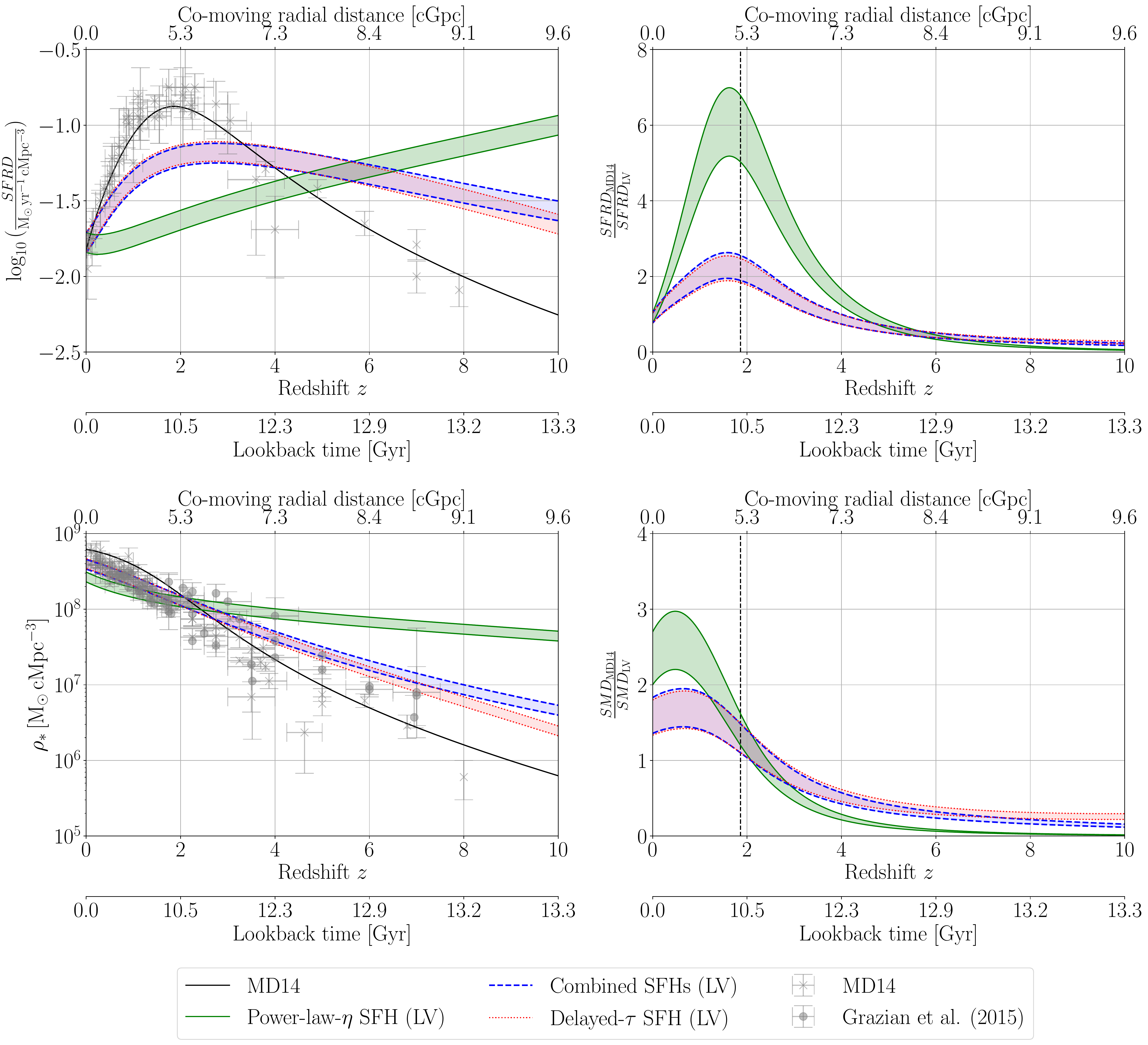}
    \caption{The SFRD (top panels) and SMD (bottom panels) over redshift and lookback time (bottom x-axis) and co-moving radial distance (top x-axis). Top left: The black solid curve shows the best-fitting function of \citetalias{Madau_Dickinson_2014} (see their eq.~15 and Eq.~\ref{eq:SFRD_Madau} in Section~\ref{sec:Introduction}) to the FUV and IR measurements (grey crosses) as shown in their figure~9 and listed in their table~1. The red dotted and green solid curves show the cosmic SFRDs as deduced from the LV galaxies assuming the delayed-$\tau$ and power-law-$\eta$ SFH model, respectively. The blue dashed line is the combined SFH model, which adopts the delayed-$\tau$ for galaxies with $M_{*} \geq 10^{10}\,\rm{M_{\odot}}$ and power-law-$\eta$ SFH model for galaxies with $M_{*}< 10^{10}\,\rm{M_{\odot}}$. The lower and upper limits of the SFRDs are given by Eq.~\ref{eq:Volume_maximum} and \ref{eq:Volume_minimum}, respectively. Top right: Ratio between the cosmic SFRD measured by \citetalias{Madau_Dickinson_2014} (their eq.~15) and the SFRD of the LV. The black dashed vertical line marks the peak of the Lilly-Madau plot at $z = 1.86$. Bottom left: SMD deduced from the measured SFRD by \citetalias{Madau_Dickinson_2014} and the SFH of the LV assuming the delayed-$\tau$ (red dotted), power-law-$\eta$ (green solid), and our combined (blue dashed) SFH models. The grey crosses and circles are observations by \citetalias{Madau_Dickinson_2014} and \citet{Grazian_2015}, who assume a recycling factor of $R=0.27$ and $R = 0.28$, respectively. These data are taken from table~2 of \citetalias{Madau_Dickinson_2014} and table~2 of \citet{Yu_2016}. The derivation of the SMDs assumes $R = 0.23$ (Section~\ref{subsec:SMD}) in order to be consistent with the analysis of \citet{Kroupa_2020} but this difference with the observations is negligible without changing the interpretation of the results. Bottom right: Ratio between the cosmic SMD deduced from \citetalias[][]{Madau_Dickinson_2014} and the LV.}
    \label{figure_SFRD}
\end{figure*}

The evolution of the cosmic SFRD and SMD derived from the LV is presented in Figure~\ref{figure_SFRD}. The LV has an SFRD of $0.014-0.019 \, \rm{M_{\odot}\,yr^{-1}\,Mpc^{-3}}$ (Section~\ref{subsubsec:Present-day SFRD}) which is fully consistent with the measured value of $0.015 \, \rm{M_{\odot}\,yr^{-1}\,Mpc^{-3}}$ at $z = 0$ by \citetalias{Madau_Dickinson_2014}. However, the evolution of the cosmic SFRD of the LV significantly differs from \citetalias{Madau_Dickinson_2014} by being systematically lower at $z \la 3$ but higher at $z \ga 5.5$ (top left panel). The delayed-$\tau$ and the combined SFH model peak globally with a value of $0.058-0.078 \, \rm{M_{\odot}\,yr^{-1}\,cMpc^{-3}}$ at $z = 2.73$ and $0.056-0.076\, \rm{M_{\odot}\,yr^{-1}\,cMpc^{-3}}$ at $z = 2.78$, respectively, while the \citetalias{Madau_Dickinson_2014} plot reaches a much higher maximum of $0.13 \, \rm{M_{\odot}\,yr^{-1}\,cMpc^{-3}}$ at $z = 1.86$. The SFRD of the power-law-$\eta$ SFH model peaks with $1.00-1.35 \, \rm{M_{\odot}\,yr^{-1}\,cMpc^{-3}}$ at $z = 18$, which corresponds to an age of the Universe of $0.2$~Gyr \citep{Planck_2016_IllustrisTNG} and the assumed onset of star formation (Section~\ref{subsubsec:Present-day averaged SFR}).
The SFRD derived from the LV is $2.16\pm0.32$ (delayed-$\tau$ SFH model), $2.22\pm0.33$ (combined SFH model), and $5.90\pm0.88$ (power-law-$\eta$ SFH model) times lower than measured by \citetalias{Madau_Dickinson_2014} (their eq.~15) at the peak of the Lilly-Madau plot located at $z = 1.86$ (top right panel). Comparing the ratio between the global and local SFRD at $z < 8$ (top left panel) yields a maximum at $z \approx 1.6$ where the SFRD derived from the LV is by a factor $2.22\pm0.33$ (delayed-$\tau$ SFH model), $2.29\pm0.34$ (combined SFH model), and $6.09\pm0.90$ (power-law-$\eta$ SFH model) times lower. 

As the SFH and the stellar mass growth of galaxies are physically connected (Section~\ref{subsec:SMD}), the discrepancy is also evident in the evolution of the local and global SMDs as shown in the bottom panels of Figure~\ref{figure_SFRD}. The SMD derived from the LV is systematically lower at $z \la 2$ and higher at $z \ga 3$ than implied by the global cosmic SFH of \citetalias{Madau_Dickinson_2014} (bottom left panel), i.e., the ratio between the cosmic SMD deduced from
\citetalias{Madau_Dickinson_2014} and the LV peaks with $1.67\pm0.25$ at $z=0.67$ (delayed-$\tau$ SFH model), $1.70\pm0.25$ at $z = 0.64$ (combined SFH model), and $2.59\pm0.38$ at $z=0.48$ (power-law-$\eta$ SFH model; bottom right panel).

The derived SMDs are compared with the observed SMD by \citetalias{Madau_Dickinson_2014} and \citet{Grazian_2015} in the bottom left panel of Figure~\ref{figure_SFRD}. As already reported by \citet{Yu_2016}, the observed SMD is inconsistent with the SFRD measured by \citetalias{Madau_Dickinson_2014} at $0.5 \la z \la 6$ \citep[see also the figures~2-4 of][]{Yu_2016}. The delayed-$\tau$ and combined SFH models agree much better with the SMD measured by \citetalias{Madau_Dickinson_2014} and \citet{Grazian_2015} at $z \la 2$ but systematically overstimate this measured SMD at higher redshifts. The power-law-$\eta$ SFH model disagrees with the observations by \citetalias{Madau_Dickinson_2014} and \citet{Grazian_2015} over the entire redshift range. 

The implications and possible interpretations of the discrepancy between the local and global SFRD and SMD are discussed in the following section.

\section{Discussion}
\label{sec:Discussion}
The physical properties of local galaxies provide the cosmological boundary conditions for galaxy evolution and structure formation models. Galaxies located in the Local Cosmological Volume \citep[][]{Karachentsev_2004,Karachentsev_2013} ought to have well-observed physical parameters like the present-day SFRs and stellar masses \citep{Karachentsev_2013_SFRproperties, Kroupa_2020}, which can be assessed to constrain their SFHs (Sections~\ref{subsec:delayedtau model}-\ref{subsec:Combined SFH model}). Assessing the updated version of the Catalogue of Neighbouring Galaxies as compiled by \cite{Karachentsev_2004,Karachentsev_2013}, \citet{Kroupa_2020} reported that the SFH of galaxies with $SFR_{0} \geq 10^{-3}\,\rm{M_{\odot}\,yr^{-1}}$ in the LV can be parametrized by a power-law of the form of $SFR(t) \propto (t - t_{\mathrm{start}})^{\eta}$ for $t_{\mathrm{start}} = 1.8\,\rm{Gyr}$ with a power-law index of $\eta = 0.18 \pm 0.03$. This implies that the majority of LV galaxies has almost constant SFRs over time being in tension with the evolution of the SFH as implied by the main sequence of star-forming galaxies \citep[figures~9 and 10 in][]{Speagle_2014}. The cosmological implications are investigated in this contribution.

\subsection{SFHs of galaxies in the LV} \label{subsec:SFHs of galaxies in the LV}
We reconstructed the SFHs of galaxies in the LV based on their present-day and present-day averaged SFRs assuming the delayed-$\tau $ (Sections~\ref{subsec:delayedtau model}) and a power-law (Section~\ref{subsec:eta model}) SFH models. For the galaxies in the LV, we found that those with $M_{*} \ga 10^{10}\,\rm{M_{\odot}}$ typically have decreasing SFHs over the last $\ga 5$~Gyr, while less mass galaxies with $SFR_{0} \ga 10^{-3}\,\rm{M_{\odot}\,yr^{-1}}$ have almost constant or even increasing SFHs. A substantial number of objects with $SFR_{0} \la 10^{-4}\,\rm{M_{\odot}\,yr^{-1}}$ have $SFR_{0} < \overline{SFR}$ suggesting a decreasing SFH (Figure~\ref{figure_Mainseqeuence}). These findings are consistent with the studies by \citet{Speagle_2014}, \citet{McGaugh_2017}, \citet{Schombert_2019}, and \citet{Kroupa_2020} according to which galaxies with $M_{*} \ga 10^{10}\,\rm{M_{\odot}}$ follow the main sequence of star-forming galaxies and less massive galaxies with $M_{*} \ga 10^{10}\,\rm{M_{\odot}}$ have a constant SFH.

For example, the MW with $M_{*} = 3.6\times10^{10}\,\rm{M_{\odot}}$ has $SFR_{0} = 1.7\,\rm{M_{\odot}\,yr^{-1}}$ and $\overline{SFR} = 3.5\,\rm{M_{\odot}\,yr^{-1}}$ resulting in a power-law index of $\eta = -0.52$ and therewith a gradually decreasing SFH over time in the power-law-$\eta$ model. The delayed-$\tau$ model yields a star formation time-scale of $\tau = 4.1$~Gyr. Thus, the MW is consistent with galaxies on the main sequence of star-forming galaxies by \citet{Speagle_2014} which have star formation time-scales in the range of $3.5 \la \tau/\rm{Gyr} \la 4.5$ \citep{Kroupa_2020}.

The LMC and M33 (Triangulum) galaxy have $M_{*} = 1.6\times10^{9}\,\rm{M_{\odot}}$ and $2.5\times10^{9}\,\rm{M_{\odot}}$ and $SFR_{0} = 0.3 \,\rm{M_{\odot}\,yr^{-1}}$ and $0.5 \,\rm{M_{\odot}\,yr^{-1}}$, respectively. Therefore, both galaxies have a very similar increasing SFH in the delayed-$\tau$ and power-law-$\eta$ SFH model (Figure~\ref{figure_SFHs}). Indeed, \citet[][]{Javadi_2017} measured the SFH across the entire disc of M33 and found that the SFR increased from $\approx 0.04 \,\rm{M_{\odot}\,yr^{-1}}$ up to $\approx 0.5 \,\rm{M_{\odot}\,yr^{-1}}$ over the last $10\,\rm{Gyr}$ with a local minimum of $\approx 0.06 \,\rm{M_{\odot}\,yr^{-1}}$ at a lookback time of $\approx 2.4 \,\rm{Gyr}$ (see their figure~6). We derived an SFR of $0.12 \,\rm{M_{\odot}\,yr^{-1}}$ (delayed-$\tau$) and $0.11\,\rm{M_{\odot}\,yr^{-1}}$ (power-law-$\eta$) at a lookback time of $10$~Gyr, which is about three times higher than measured by \citet{Javadi_2017}. A possible explanation for these differences could be that the SFH of M33 is likely affected by interactions with M31. Such dynamical effects have not been taking account by our SFH models. Similarly, the Magellanic Clouds have an SFH which is influenced by their mutual gravitational interaction \citep[][]{Weisz_2013} and encounters with the MW. In particular, the SFH of the MCs increased over the last $3.5\,\rm{Gyr}$ (figure~5 of \citealt{Weisz_2013} and figure~2 of \citealt{Massana_2022}), which is likely caused by star formation triggering during interactions between the MCs-MW system.

\subsection{Caveats of the adopted SFH parametrization and outlooks} \label{subsec:CaveatsofSFHparameterization}
The parametrization of the galactic SFHs by the delayed-$\tau$ (Section~\ref{subsec:delayedtau model}) and power-law (Section~\ref{subsec:eta model}) model come along with some caveats as discussed in the following. The SFRs given by the power-law SFH model diverge to infinity at $t = t_{\mathrm{start}}$ for a power-law index of $\eta < 0$ which fails in describing the rising part of the SFH at early galaxy evolution \citep{Reddy_2012, Carnall_2019}. This in turn affects the general slope of the SFHs  (Figure~\ref{figure_SFHs}) and yields too high SFRDs and SMDs at redshifts $z \ga 6$ (see the left-hand panels of Figure~\ref{figure_SFRD}) being in disagreement with the observations. 

The delayed-$\tau$ SFH model accounts for rising star formation during early stages of galaxy formation but also allows for the description of gradually increasing SFHs by allowing negative values of $\tau$ (Equation~\ref{eq:delayedtau}) providing therewith a better description of galaxy evolution. Consequently, this parametrization yields to lower cosmic SFRDs and SMDs compared to the power-law model. According to Figure~\ref{figure_Mainseqeuence}, the majority of star-forming LV galaxies with $SFR_{0} \geq 10^{-3}\,\rm{M_{\odot}\,yr^{-1}}$ and $M_{*} \la 10^{10}\,\rm{M_{\odot}}$ have an almost constant SFH over a time-scale of $13.6$~Gyr (see also \citealt{Kroupa_2020} for a more extended discussion). The combined SFH model (Section~\ref{subsec:Combined SFH model}) broadly accounts for that by adopting the delayed-$\tau$ model for galaxies with $M_{*} \geq 10^{10}\,\rm{M_{\odot}}$ and power-law-$\eta$ model for galaxies with $M_{*} < 10^{10}\,\rm{M_{\odot}}$. The cosmic SFRDs and SMDs derived from the combined model only slightly differs from the delayed-$\tau$ model as the low-mass galaxies make only a small contribution to the SFRDs. More observables e.g. metallicity and chemical composition to fit the model parameters would be required. Unfortunately, the metallicities of the LV galaxies are not listed in the Catalogue of Neighbouring Galaxies \citep{Karachentsev_2004, Karachentsev_2013}. Chemical galaxy evolution models \citep[e.g.][]{Yan_2019b, Gjergo_2023} along with metallicity measurements would pin down further the SFH of LV galaxies.  

Another method to reconstruct the evolution of the local SFRD would be to derive the SFHs of LV galaxies using colour magnitude diagrams as performed e.g. by \citet{Weisz_2014} for Local Group (LG) dwarfs. Such analysis would be required to affirm the tension between the local and global cosmic SFRDs.

Some studies argued that the cosmic SFRD can be described by a lognormal SFH  \citep[e.g.][]{Behroozi_2013, Gladders_2013, Abramson_2016}  which is typically parametrized by a normalization constant, the logarithmic delay time, and the star formation time-scale. Because of these three unknown parameters, the lognormal SFH of LV galaxies cannot be just reconstructed by their present-day SFRs and stellar masses as presented in this study. In fact, \citet{Carnall_2019} parametrized the SFHs of galaxies in the spectroscopic redshift range
$0.05<z<0.08$ of the Galaxy and Mass
Assembly (GAMA) Survey \citep{Driver_2009, Driver_2016} of DR3 \citep{Baldry_2018}. Interestingly, they showed that the derived SFRDs assuming the exponentially declining and delayed-$\tau$ but also the lognormal and double power-law SFH models (see their section~2) are inconsistent with the Lilly-Madau plot of \citetalias{Madau_Dickinson_2014}. We discuss this further in the following section.

\subsection{Constraints on the matter homogeneity} \label{subsec:Cosmic SFRD of the LV}
The LV has an SFRD of $0.014-0.019\, \rm{M_{\odot}\,yr^{-1}\,Mpc^{-3}}$ (Section~\ref{subsubsec:Present-day SFRD}) being therewith in agreement with the measurements by \citetalias{Madau_Dickinson_2014}. Calculating the SFRD of the LG environment by selecting only galaxies with $D < 1.5\,\rm{Mpc}$ gives a much higher value of $0.22-0.29 \, \rm{M_{\odot}\,yr^{-1}\,Mpc^{-3}}$. This is because the LG (i.e. the MW and M31 galaxies with their surrounding dwarf galaxies) forms an overdensity within the LV as evident e.g. in the figures~1 and 2 of \citet{Karachentsev_2018}.

\citet{Kroupa_2020} reported that the majority of galaxies located in the LV has a constant SFH over a time-scale of $t_{\mathrm{sf}} = 12$~Gyr. As the global co-moving SFRD increases from the present time up to the Lilly-Madau peak at $z = 1.86$ by a factor of $8.9$ \citepalias{Madau_Dickinson_2014}, the findings by \citet{Kroupa_2020} raise the question if the SFH of the LV is consistent with the global SFH of the Universe. Supplementing their analysis, we deduce the evolution of the SFRD by reconstructing the cosmic SFH of LV galaxies. We found that the SFHs implied from the LV galaxies systematically underestimate the SFRDs at $z \la 3$ and overestimates the SFRDs at $z \ga 6$ as in comparison to those of \citetalias{Madau_Dickinson_2014}. In particular, the SFRD at the peak of the Lilly-Madau plot is $2.16\pm0.32$ and $5.90\pm0.88$ times lower in the case of the delayed-$\tau$ and power-law-$\eta$ SFH model, respectively. The combined SFH model underestimates the SFRD by a factor of $2.22\pm0.33$ at $z = 1.86$. Thus, the SFHs of galaxies located in the LV disagree with the expectations from the Lilly-Madau plot.

Similar conclusions have been reached by \citet{Carnall_2019} who parametrized the SFH of galaxies in the redshift range of $0.05<z<0.08$ from the GAMA Survey (Section~\ref{subsec:CaveatsofSFHparameterization}). They derived a peak of the SFRD with a value of $\approx 0.05\,\rm{M_{\odot}\,yr^{-1}\,cMpc^{-3}}$ at $z \approx 0.4$ (see their figure~10). While the maximum value is similar to our results, the peak appeared at much later times. In a companion study, \citet{Leja_2019} applied non-parametric SFH models, which disagrees with the shape of the cosmic SFRD evolution suggested by \citetalias{Madau_Dickinson_2014} (left-hand panel of figure~13).

The SFH and the stellar mass growth are physically connected (Section~\ref{subsec:SMD}) such that the disagreement between the local and global SFHs is also evident in the cosmic SMDs deduced from the SFRDs. The SMD derived from the local SFHs is up to $1.67\pm0.25$ (delayed-$\tau$ SFH model), $1.70\pm0.25$ (combined SFH model), and $2.59\pm0.39$ (power-law-$\eta$ SFH model) times lower at $z \la 2$ and systematically higher at $z \ga 3$ than implied by \citetalias{Madau_Dickinson_2014}. Interestingly, several studies reported an inconsistency between the observed SMD and SFRD over cosmic time \citep[e.g.][]{Hopkins_2006,Hopkins_2008,Wilkins_2008,Madau_Dickinson_2014,Grazian_2015,Leja_2015,Tomczak_2016,Yu_2016}. For example, \citet{Yu_2016} found that the observed SFRD by \citetalias{Madau_Dickinson_2014} is by a factor of about 2 higher than the SFH derived from the observed SMD \citep{Madau_Dickinson_2014,Grazian_2015}. The reason is not understood yet but it has been proposed that a top-heavy or bottom-light IMF could alleviate or even resolve the discrepancy \citep{Yu_2016}. The affect of a varying IMF on the Lilly-Madau plot in the context of the integrated-galactic IMF \citep[IGIMF, ][]{Kroupa_2003} has been studied by \citet{Chruslinska_2020} but also a correction of the SMD would be required for a direct comparison. Investigating the effect of a varying IMF on the present-day and present-day averaged SFR would be valuable but is beyond the scope of this work. 

The SMD derived from the LV assuming the delayed-$\tau$ and combined SFH model agrees much better with the observed SMD at $z \la 2$ compared with the power-law model but is still systematically higher at $z \ga 2$. Upcoming observations of high-redshift galaxies by the James Webb Space Telescope (JWST) could potentially increase the observed SMD at $z \ga 5$ alleviating therewith this discrepancy. 

The tension between the local and global SFHs could be caused by several reasons. First of all, the measurements of the LV could have a systematic error by overestimating the present-day SFRs or underestimating the stellar masses \citep[see also section~4 of][]{Kroupa_2020}. This seems to be unlikely because the spatial proximity of the LV to the MW should provide access to reliable and model-independent galaxy data and \citet{Carnall_2019} also derived lower SFRDs than \citetalias{Madau_Dickinson_2014} based on the GAMA Survey. Alternatively, the SFRD given by \citet[][]{Madau_Dickinson_2014} could be systematically overestimated. However, the measured SFRDs especially at high redshifts of $z \ga 1$ should rather be treated as lower instead of upper limits because of the potential non-detection of low-surface-brightness galaxies \citep[][]{Kim_2022} and high-redshift galaxies \citep[e.g.][]{Castellano_2023,Helton_2023}. 

Secondly, the assumed parametrization of the applied SFH models may be inadequate to describe galaxy evolution. \citet[][]{Speagle_2014} showed that the delayed-$\tau$ model provides an accurate description of the SFHs of galaxies, which strictly follow the main sequence of star-forming galaxies (see their figures~9 and 10). \citet[][]{Kroupa_2020} argued that the power-law-$\eta$ model parametrizes the SFH of LV galaxies (see their section~3.3). Therefore, the combined SFH model should provide a reasonable parametrization of galaxy evolution. Modelling the SFH of all galaxies with the power-law-$\eta$ model conflicts with the observed SMD (see the bottom left panel of \ref{figure_SFRD}). 

Thirdly, the LV is not a representative sample of the observed Universe as discussed in section~4 of \citet{Kroupa_2020}. As pointed out by \citet[][]{Peebles_2010}, the spatial arrangement of major galaxies in a sphere with $D~<~8$~Mpc disagrees with the hierarchical structure formation of the $\Lambda$CDM framework. Furthermore, the LV is embedded in a local void, which is evident across the entire electromagnetic spectrum ranging from the radio to the X-ray regime \citep[e.g.][]{Maddox_1990,Zucca_1997,Karachentsev_2012,Keenan_2013,Rubart_Schwarz_2013, Rubart_2014, Boehringer_2015,Karachentsev_2018,Boehringer_2020,Wong_2022}. For example, \citet{Karachentsev_2012} measured an average total matter density of $\Omega_{\mathrm{m,loc}} = 0.08 \pm 0.02$ for a sphere with $50$~Mpc centred around the MW, which is about four times lower than the global value given by \citet[][]{Planck_2016_IllustrisTNG}. The local Universe is underdense on even much larger scales. \citet*{Keenan_2013} found that the matter density is about two times lower than the cosmic mean density on a $300$~cMpc radial scale. The density contrast increases between $300$~cMpc and $600-800$~cMpc by a factor of about 2 (see their figure~11). This so-called KBC void contradicts the $\Lambda$CDM model at more than $5\sigma$ confidence \citep[][]{Haslbauer_2020}, which predicts root-mean-square (rms) density fluctuations of only $3.2$~per cent on a 300~Mpc scale. The evolution of galaxies located in underdense regions like the LV may be retarded compared to galaxies in overdensities \citep[][]{Douglass_2017}. 

Finally, the Lilly-Madau plot as quantitatively described by Eq.~\ref{eq:SFRD_Madau} is sensitive not only to SFRs but also to the number density of galaxies. Thus, the here reported tension could imply that the peak of the Lilly-Madau plot is mainly caused by an overdensity at $z \approx 1.86$, which corresponds to a co-moving radius distance of $\approx 5.1 \, \rm{cGpc}$. Interestingly, the evolution of the co-moving mass density of molecular gas in galaxies partially follows the shape of the Lilly-Madau plot \citep[e.g.][]{Decarli_2016,Riechers_2019, Walter_2020}. As star formation is mainly driven by molecular gas, the Lilly-Madau plot could be therefore the consequence of the underlying molecular gas density. This would be consistent with an overdensity at $z \approx 2$ since the matter field should also be evident in the gas component. 

Although this leads to an entirely new interpretation of the Lilly-Madau peak, indications of a significant overdensity on very large scales have been reported by several studies \citep[e.g.][]{Horvarth_2014,Horvarth_2015,Horvath_2020,Migkas_2021,Secrest_2022}. Inhomogeneities in the matter field affect the measurements of cosmological parameters as they change the motion of galaxies \citep[][]{Haslbauer_2020}. Remarkably, there is observationally evidence that $H_{0}$ decreases up to $z \approx 1.7$ in the SN data \citep[figures~2 and 3 of][but see also \citealt{Krishnan_2020} and \citealt{Dainotti_2021}]{Jia_2023}. Measuring $H_{0}$ from six gravitationally lensed quasars with lens redshifts between $z = 0.295$ and $0.745$, \citet{Wong_2020} obtained that $H_{0}$ decreases with an increasing lens redshift but only with a significance of $1.9\sigma$ (see their figure~A1). In this scenario, the Hubble tension is not only caused by the 300~Mpc scale KBC void \citep[][]{Haslbauer_2020} but is rather the consequence of an $\approx 5.1$~cGpc scale inhomogeneity.

Further evidence for large-scale matter flows and inhomogeneities comes from \citet[][]{Migkas_2021}, who reported bulk flows with velocities of $\approx 900\,\rm{km\,s^{-1}}$ to at least $500$~cMpc scales from galaxy cluster data. \citet[][]{Secrest_2022} found that the dipole amplitude and direction of distant radio galaxies and quasars significantly differ from the cosmic microwave background (CMB) dipole violating the cosmological principle with a significance of $5.1\sigma$. The authors argue that this favours the existence of an overdensity on large scales at Galactic coordinates of $(l, b) = (217^{\circ}\pm10^{\circ}, +20^{\circ}\pm7^{\circ})$. Interestingly, the Hercules–Corona Borealis Great Wall as observed by an excess of gamma-ray bursts (GRBs) between $z = 1.6$ and $2.1$ \citep{Horvarth_2014,Horvarth_2015,Horvath_2020} coincides with the Lilly-Madau peak at $z = 1.86$. However, these GRBs are observed at the second, third, and fourth Galactic quadrants and not towards $(l, b) = (217^{\circ} \pm 10^{\circ}, +20^{\circ} \pm 7^{\circ})$.

Large-scale anisotropies in the morphological distribution of galaxies have been studied by \citet{Javanmardi_2017} who showed that the Northern hemisphere has more late-type galaxies, while the Southern hemisphere has more early-type galaxies. Notably, the Southern ecliptic hemisphere has also more power in the CMB than the Northern ecliptic hemisphere \citep{Eriksen_2004,Schwarz_2016}. 

A large-scale density variation could potentially resolve the so-called $\sigma_{8}$ problem \citep[see also][]{Adil_2023}, which states that the present rms matter fluctuation averaged over a sphere with a radius of $8\,h^{-1}$~Mpc is locally lower \citep[$\sigma_{8} =0.785\pm0.029$;][]{Abbott_2022} than on global scales \citep[$\sigma_{8} = 0.8159\pm0.0086$;][]{Planck_2016_IllustrisTNG}. This is because density fluctuations within a larger scale underdensity could be suppressed making the Universe to appear locally ``less clumpy" than predicted by the $\Lambda$CDM model.

The accelerated expansion of the Universe could also be an apparent effect due to large-scale inhomogeneities that grow with cosmic time \citep{Wiltshire_2007a,Wiltshire_2007b,Wiltshire_2009,Wiltshire_2019}.

More precise observations of the matter distribution are required to test the here proposed hypothesis that the peak of the Lilly-Madau peak could be the imprint of a large-scale density variation. Directional anisotropies in the Lilly-Madau plot could be caused by different density contrasts at different sky directions. Gpc-large-scale matter inhomogeneities would violate the cosmological principle of the $\Lambda$CDM model, which predicts an upper limit to the scale of homogeneity of about $260\,h^{-1} \, \rm{Mpc}$, being homogeneous on larger scales \citep{Yadav_2010}. However, several observations indeed question the scale of homogeneity of the $\Lambda$CDM model as outlined above \citep[e.g.][]{Gott_2005, Clowes_2012, Keenan_2013, Horvarth_2014,Javanmardi_2015,Javanmardi_2017,Colin_2019,Haslbauer_2020,Migkas_2020,Migkas_2021, Secrest_2021, Secrest_2022, KumarAluri_2023}. Interestingly, Milgromian  \citep[MOND;][]{Milgrom_1983} cosmological models with a non-zero cosmological constant also predict large-scale homogeneity on a scale of several hundred Mpc \citep[][]{Sanders_1998}. 

\section{Conclusion} \label{sec:Conclusion}
Supplementing the analysis of \citet{Kroupa_2020}, we investigated the cosmological implications of the SFHs of galaxies located in the LV. We reconstructed the SFHs with the delayed-$\tau$ (Section~\ref{subsec:delayedtau model}) and power-law-$\eta$ (Section~\ref{subsec:eta model}) models in order to calculate the SFRD of the LV over cosmic time. The near-constancy of the SFH of LV galaxies as reported by \citet{Kroupa_2020} has implications for the Lilly-Madau plot, which shows that the SFRD increases from present time up to redshift $z = 1.86$ by a factor of about 9. While the local SFRD is consistent with \citetalias[][]{Madau_Dickinson_2014}, the derived SFHs parametrized by the delayed-$\tau$ and power-law-$\eta$ SFH models underestimate the cosmic SFRD at $z \la 3$ and overestimate the SFRD at $z \ga 6$. In particular, the SFRD is   $2.16\pm0.32$ and $5.90\pm0.88$ times lower than measured at the peak of the Lilly-Madau plot at $z=1.86$ if the delayed-$\tau$ and power-law-$\eta$ SFH models are assumed, respectively. Combining both SFH models by adopting the delayed-$\tau$ model for galaxies with $M_{*} \geq 10^{10}\,\rm{M_{\odot}}$ and the power-law-$\eta$ model for galaxies with $M_{*} < 10^{10}\,\rm{M_{\odot}}$ (Section~\ref{subsec:Combined SFH model}) underestimates the SFRD by a factor of 
$2.22\pm0.33$ at $z = 1.86$. Several reasons for the disagreement between the local and global cosmic SFRDs such as measurement biases or parametrization of the SFHs have been discussed in Section~\ref{sec:Discussion}.

In conclusion, we found that the derived SFHs of LV galaxies are in disagreement with the Lilly-Madau plot (see also \citealt{Carnall_2019} and \citealt{Leja_2019} for the GAMA survey). As the cosmological SFH of the local Universe sets constraints on the matter homogeneity, the discrepancy could be caused by the local underdensity \citep[][]{Karachentsev_2012,Karachentsev_2018} and/or the peak of the Lilly-Madau plot is the imprint of $\approx~5$~cGpc-scale density variations at $z \approx 1.86$. Independent evidence of significant inhomogeneity on $>100$~Mpc scales is discussed in Section~\ref{subsec:Cosmic SFRD of the LV}. Precise measurements of the SFH and the number counts of galaxies would be required to affirm such an interpretation of the Lilly-Madau peak. 

\section*{Data availability}
The data underlying this article are available in the article. The physical parameters of galaxies located in the LV are taken from the updated version of the Catalogue of Neighbouring Galaxies \citep{Karachentsev_2004, Karachentsev_2013} as explained in Section~\ref{subsec:Data}. 

\section*{Acknowledgements}
The authors thank an anonymous referee for helpful suggestions improving the manuscript. Moritz Haslbauer acknowledges support from the European Southern Observatory (ESO) Early-Career Visitor Programme. The project has been discussed during the ``Jind\v{r}ich\r{u}v Hradec-meeting on stellar populations, gravitational dynamics and Milgromian-based cosmology" held at the Florian Palace in Jind\v{r}ich\r{u}v Hradec (Czech Republic). The authors acknowledge the Deutscher Akademischer Austauschdienst (DAAD)-Eastern-European exchange programme for financial support and the Florian Palace for hosting this meeting.

\bibliographystyle{mnras}
\bibliography{Madau}

\begin{thebibliography}{}
\makeatletter
\relax
\def\mn@urlcharsother{\let\do\@makeother \do\$\do\&\do\#\do\^\do\_\do\%\do\~}
\def\mn@doi{\begingroup\mn@urlcharsother \@ifnextchar [ {\mn@doi@}
  {\mn@doi@[]}}
\def\mn@doi@[#1]#2{\def\@tempa{#1}\ifx\@tempa\@empty \href
  {http://dx.doi.org/#2} {doi:#2}\else \href {http://dx.doi.org/#2} {#1}\fi
  \endgroup}
\def\mn@eprint#1#2{\mn@eprint@#1:#2::\@nil}
\def\mn@eprint@arXiv#1{\href {http://arxiv.org/abs/#1} {{\tt arXiv:#1}}}
\def\mn@eprint@dblp#1{\href {http://dblp.uni-trier.de/rec/bibtex/#1.xml}
  {dblp:#1}}
\def\mn@eprint@#1:#2:#3:#4\@nil{\def\@tempa {#1}\def\@tempb {#2}\def\@tempc
  {#3}\ifx \@tempc \@empty \let \@tempc \@tempb \let \@tempb \@tempa \fi \ifx
  \@tempb \@empty \def\@tempb {arXiv}\fi \@ifundefined
  {mn@eprint@\@tempb}{\@tempb:\@tempc}{\expandafter \expandafter \csname
  mn@eprint@\@tempb\endcsname \expandafter{\@tempc}}}

\bibitem[\protect\citeauthoryear{{Abbott} et~al.,}{{Abbott}
  et~al.}{2023}]{Abbott_2022}
{Abbott} T.~M.~C.,  et~al., 2023, \mn@doi [\prd] {10.1103/PhysRevD.107.023531},
  \href {https://ui.adsabs.harvard.edu/abs/2023PhRvD.107b3531A} {107, 023531}

\bibitem[\protect\citeauthoryear{{Abramson}, {Gladders}, {Dressler}, {Oemler},
  {Poggianti}  \& {Vulcani}}{{Abramson} et~al.}{2016}]{Abramson_2016}
{Abramson} L.~E.,  {Gladders} M.~D.,  {Dressler} A.,  {Oemler} A. J.,
  {Poggianti} B.,   {Vulcani} B.,  2016, \mn@doi [\apj]
  {10.3847/0004-637X/832/1/7}, \href
  {https://ui.adsabs.harvard.edu/abs/2016ApJ...832....7A} {832, 7}

\bibitem[\protect\citeauthoryear{{Adil}, {Akarsu}, {Malekjani}, {Colg{\'a}in},
  {Pourojaghi}, {Sen}  \& {Sheikh-Jabbari}}{{Adil} et~al.}{2023}]{Adil_2023}
{Adil} S.~A.,  {Akarsu} {\"O}.,  {Malekjani} M.,  {Colg{\'a}in} E.~{\'O}.,
  {Pourojaghi} S.,  {Sen} A.~A.,   {Sheikh-Jabbari} M.~M.,  2023, \mn@doi
  [arXiv e-prints] {10.48550/arXiv.2303.06928}, \href
  {https://ui.adsabs.harvard.edu/abs/2023arXiv230306928A} {p. arXiv:2303.06928}

\bibitem[\protect\citeauthoryear{{Aluri} et~al.,}{{Aluri}
  et~al.}{2023}]{KumarAluri_2023}
{Aluri} P.~K.,  et~al., 2023, \mn@doi [Classical and Quantum Gravity]
  {10.1088/1361-6382/acbefc}, \href
  {https://ui.adsabs.harvard.edu/abs/2023CQGra..40i4001K} {40, 094001}

\bibitem[\protect\citeauthoryear{{Baldry} et~al.,}{{Baldry}
  et~al.}{2018}]{Baldry_2018}
{Baldry} I.~K.,  et~al., 2018, \mn@doi [\mnras] {10.1093/mnras/stx3042}, \href
  {https://ui.adsabs.harvard.edu/abs/2018MNRAS.474.3875B} {474, 3875}

\bibitem[\protect\citeauthoryear{{Baumgardt} \& {Makino}}{{Baumgardt} \&
  {Makino}}{2003}]{Baumgardt_2003}
{Baumgardt} H.,  {Makino} J.,  2003, \mn@doi [\mnras]
  {10.1046/j.1365-8711.2003.06286.x}, \href
  {https://ui.adsabs.harvard.edu/abs/2003MNRAS.340..227B} {340, 227}

\bibitem[\protect\citeauthoryear{{Behroozi}, {Wechsler}  \&
  {Conroy}}{{Behroozi} et~al.}{2013}]{Behroozi_2013}
{Behroozi} P.~S.,  {Wechsler} R.~H.,   {Conroy} C.,  2013, \mn@doi [\apj]
  {10.1088/0004-637X/770/1/57}, \href
  {https://ui.adsabs.harvard.edu/abs/2013ApJ...770...57B} {770, 57}

\bibitem[\protect\citeauthoryear{{B{\"o}hringer}, {Chon}, {Bristow}  \&
  {Collins}}{{B{\"o}hringer} et~al.}{2015}]{Boehringer_2015}
{B{\"o}hringer} H.,  {Chon} G.,  {Bristow} M.,   {Collins} C.~A.,  2015,
  \mn@doi [\aap] {10.1051/0004-6361/201424817}, \href
  {https://ui.adsabs.harvard.edu/abs/2015A&A...574A..26B} {574, A26}

\bibitem[\protect\citeauthoryear{{B{\"o}hringer}, {Chon}  \&
  {Collins}}{{B{\"o}hringer} et~al.}{2020}]{Boehringer_2020}
{B{\"o}hringer} H.,  {Chon} G.,   {Collins} C.~A.,  2020, \mn@doi [\aap]
  {10.1051/0004-6361/201936400}, \href
  {https://ui.adsabs.harvard.edu/abs/2020A&A...633A..19B} {633, A19}

\bibitem[\protect\citeauthoryear{{Carnall}, {Leja}, {Johnson}, {McLure},
  {Dunlop}  \& {Conroy}}{{Carnall} et~al.}{2019}]{Carnall_2019}
{Carnall} A.~C.,  {Leja} J.,  {Johnson} B.~D.,  {McLure} R.~J.,  {Dunlop}
  J.~S.,   {Conroy} C.,  2019, \mn@doi [\apj] {10.3847/1538-4357/ab04a2}, \href
  {https://ui.adsabs.harvard.edu/abs/2019ApJ...873...44C} {873, 44}

\bibitem[\protect\citeauthoryear{{Castellano} et~al.,}{{Castellano}
  et~al.}{2023}]{Castellano_2023}
{Castellano} M.,  et~al., 2023, \mn@doi [\apjl] {10.3847/2041-8213/accea5},
  \href {https://ui.adsabs.harvard.edu/abs/2023ApJ...948L..14C} {948, L14}

\bibitem[\protect\citeauthoryear{{Chru{\'s}li{\'n}ska},
  {Je{\v{r}}{\'a}bkov{\'a}}, {Nelemans}  \& {Yan}}{{Chru{\'s}li{\'n}ska}
  et~al.}{2020}]{Chruslinska_2020}
{Chru{\'s}li{\'n}ska} M.,  {Je{\v{r}}{\'a}bkov{\'a}} T.,  {Nelemans} G.,
  {Yan} Z.,  2020, \mn@doi [\aap] {10.1051/0004-6361/202037688}, \href
  {https://ui.adsabs.harvard.edu/abs/2020A&A...636A..10C} {636, A10}

\bibitem[\protect\citeauthoryear{{Clowes}, {Campusano}, {Graham}  \&
  {S{\"o}chting}}{{Clowes} et~al.}{2012}]{Clowes_2012}
{Clowes} R.~G.,  {Campusano} L.~E.,  {Graham} M.~J.,   {S{\"o}chting} I.~K.,
  2012, \mn@doi [\mnras] {10.1111/j.1365-2966.2011.19719.x}, \href
  {https://ui.adsabs.harvard.edu/abs/2012MNRAS.419..556C} {419, 556}

\bibitem[\protect\citeauthoryear{{Cole} et~al.,}{{Cole}
  et~al.}{2001}]{Cole_2001}
{Cole} S.,  et~al., 2001, \mn@doi [\mnras] {10.1046/j.1365-8711.2001.04591.x},
  \href {https://ui.adsabs.harvard.edu/abs/2001MNRAS.326..255C} {326, 255}

\bibitem[\protect\citeauthoryear{{Colin}, {Mohayaee}, {Rameez}  \&
  {Sarkar}}{{Colin} et~al.}{2019}]{Colin_2019}
{Colin} J.,  {Mohayaee} R.,  {Rameez} M.,   {Sarkar} S.,  2019, \mn@doi [\aap]
  {10.1051/0004-6361/201936373}, \href
  {https://ui.adsabs.harvard.edu/abs/2019A&A...631L..13C} {631, L13}

\bibitem[\protect\citeauthoryear{{Cook} et~al.,}{{Cook}
  et~al.}{2014}]{Cook_2014}
{Cook} D.~O.,  et~al., 2014, \mn@doi [\mnras] {10.1093/mnras/stu1787}, \href
  {https://ui.adsabs.harvard.edu/abs/2014MNRAS.445..899C} {445, 899}

\bibitem[\protect\citeauthoryear{{Dainotti}, {De Simone}, {Schiavone},
  {Montani}, {Rinaldi}  \& {Lambiase}}{{Dainotti} et~al.}{2021}]{Dainotti_2021}
{Dainotti} M.~G.,  {De Simone} B.,  {Schiavone} T.,  {Montani} G.,  {Rinaldi}
  E.,   {Lambiase} G.,  2021, \mn@doi [\apj] {10.3847/1538-4357/abeb73}, \href
  {https://ui.adsabs.harvard.edu/abs/2021ApJ...912..150D} {912, 150}

\bibitem[\protect\citeauthoryear{{Decarli} et~al.,}{{Decarli}
  et~al.}{2016}]{Decarli_2016}
{Decarli} R.,  et~al., 2016, \mn@doi [\apj] {10.3847/1538-4357/833/1/69}, \href
  {https://ui.adsabs.harvard.edu/abs/2016ApJ...833...69D} {833, 69}

\bibitem[\protect\citeauthoryear{{Douglass} \& {Vogeley}}{{Douglass} \&
  {Vogeley}}{2017}]{Douglass_2017}
{Douglass} K.~A.,  {Vogeley} M.~S.,  2017, \mn@doi [\apj]
  {10.3847/1538-4357/aa5e53}, \href
  {https://ui.adsabs.harvard.edu/abs/2017ApJ...837...42D} {837, 42}

\bibitem[\protect\citeauthoryear{{Driver} et~al.,}{{Driver}
  et~al.}{2009}]{Driver_2009}
{Driver} S.~P.,  et~al., 2009, \mn@doi [Astronomy and Geophysics]
  {10.1111/j.1468-4004.2009.50512.x}, \href
  {https://ui.adsabs.harvard.edu/abs/2009A&G....50e..12D} {50, 5.12}

\bibitem[\protect\citeauthoryear{{Driver} et~al.,}{{Driver}
  et~al.}{2016}]{Driver_2016}
{Driver} S.~P.,  et~al., 2016, \mn@doi [\mnras] {10.1093/mnras/stv2505}, \href
  {https://ui.adsabs.harvard.edu/abs/2016MNRAS.455.3911D} {455, 3911}

\bibitem[\protect\citeauthoryear{{Eriksen}, {Hansen}, {Banday}, {G{\'o}rski}
  \& {Lilje}}{{Eriksen} et~al.}{2004}]{Eriksen_2004}
{Eriksen} H.~K.,  {Hansen} F.~K.,  {Banday} A.~J.,  {G{\'o}rski} K.~M.,
  {Lilje} P.~B.,  2004, \mn@doi [\apj] {10.1086/382267}, \href
  {https://ui.adsabs.harvard.edu/abs/2004ApJ...605...14E} {605, 14}

\bibitem[\protect\citeauthoryear{{Gjergo} et~al.,}{{Gjergo}
  et~al.}{2023}]{Gjergo_2023}
{Gjergo} E.,  et~al., 2023, \mn@doi [\apjs] {10.3847/1538-4365/aca7c7}, \href
  {https://ui.adsabs.harvard.edu/abs/2023ApJS..264...44G} {264, 44}

\bibitem[\protect\citeauthoryear{{Gladders}, {Oemler}, {Dressler}, {Poggianti},
  {Vulcani}  \& {Abramson}}{{Gladders} et~al.}{2013}]{Gladders_2013}
{Gladders} M.~D.,  {Oemler} A.,  {Dressler} A.,  {Poggianti} B.,  {Vulcani} B.,
    {Abramson} L.,  2013, \mn@doi [\apj] {10.1088/0004-637X/770/1/64}, \href
  {https://ui.adsabs.harvard.edu/abs/2013ApJ...770...64G} {770, 64}

\bibitem[\protect\citeauthoryear{{Gott}, {Juri{\'c}}, {Schlegel}, {Hoyle},
  {Vogeley}, {Tegmark}, {Bahcall}  \& {Brinkmann}}{{Gott}
  et~al.}{2005}]{Gott_2005}
{Gott} J.~Richard I.,  {Juri{\'c}} M.,  {Schlegel} D.,  {Hoyle} F.,  {Vogeley}
  M.,  {Tegmark} M.,  {Bahcall} N.,   {Brinkmann} J.,  2005, \mn@doi [\apj]
  {10.1086/428890}, \href
  {https://ui.adsabs.harvard.edu/abs/2005ApJ...624..463G} {624, 463}

\bibitem[\protect\citeauthoryear{{Grazian} et~al.,}{{Grazian}
  et~al.}{2015}]{Grazian_2015}
{Grazian} A.,  et~al., 2015, \mn@doi [\aap] {10.1051/0004-6361/201424750},
  \href {https://ui.adsabs.harvard.edu/abs/2015A&A...575A..96G} {575, A96}

\bibitem[\protect\citeauthoryear{{Haslbauer}, {Banik}  \& {Kroupa}}{{Haslbauer}
  et~al.}{2020}]{Haslbauer_2020}
{Haslbauer} M.,  {Banik} I.,   {Kroupa} P.,  2020, \mn@doi [\mnras]
  {10.1093/mnras/staa2348}, \href
  {https://ui.adsabs.harvard.edu/abs/2020MNRAS.499.2845H} {499, 2845}

\bibitem[\protect\citeauthoryear{{Helton} et~al.,}{{Helton}
  et~al.}{2023}]{Helton_2023}
{Helton} J.~M.,  et~al., 2023, arXiv e-prints, \href
  {https://ui.adsabs.harvard.edu/abs/2023arXiv230210217H} {p. arXiv:2302.10217}

\bibitem[\protect\citeauthoryear{{Hopkins} \& {Beacom}}{{Hopkins} \&
  {Beacom}}{2006}]{Hopkins_2006}
{Hopkins} A.~M.,  {Beacom} J.~F.,  2006, \mn@doi [\apj] {10.1086/506610}, \href
  {https://ui.adsabs.harvard.edu/abs/2006ApJ...651..142H} {651, 142}

\bibitem[\protect\citeauthoryear{{Hopkins} \& {Beacom}}{{Hopkins} \&
  {Beacom}}{2008}]{Hopkins_2008}
{Hopkins} A.~M.,  {Beacom} J.~F.,  2008, \mn@doi [\apj] {10.1086/589809}, \href
  {https://ui.adsabs.harvard.edu/abs/2008ApJ...682.1486H} {682, 1486}

\bibitem[\protect\citeauthoryear{{Horv{\'a}th}, {Hakkila}  \&
  {Bagoly}}{{Horv{\'a}th} et~al.}{2014}]{Horvarth_2014}
{Horv{\'a}th} I.,  {Hakkila} J.,   {Bagoly} Z.,  2014, \mn@doi [\aap]
  {10.1051/0004-6361/201323020}, \href
  {https://ui.adsabs.harvard.edu/abs/2014A&A...561L..12H} {561, L12}

\bibitem[\protect\citeauthoryear{{Horv{\'a}th}, {Bagoly}, {Hakkila}  \&
  {T{\'o}th}}{{Horv{\'a}th} et~al.}{2015}]{Horvarth_2015}
{Horv{\'a}th} I.,  {Bagoly} Z.,  {Hakkila} J.,   {T{\'o}th} L.~V.,  2015,
  \mn@doi [\aap] {10.1051/0004-6361/201424829}, \href
  {https://ui.adsabs.harvard.edu/abs/2015A&A...584A..48H} {584, A48}

\bibitem[\protect\citeauthoryear{{Horvath}, {Sz{\'e}csi}, {Hakkila},
  {Szab{\'o}}, {Racz}, {T{\'o}th}, {Pinter}  \& {Bagoly}}{{Horvath}
  et~al.}{2020}]{Horvath_2020}
{Horvath} I.,  {Sz{\'e}csi} D.,  {Hakkila} J.,  {Szab{\'o}} {\'A}.,  {Racz}
  I.~I.,  {T{\'o}th} L.~V.,  {Pinter} S.,   {Bagoly} Z.,  2020, \mn@doi
  [\mnras] {10.1093/mnras/staa2460}, \href
  {https://ui.adsabs.harvard.edu/abs/2020MNRAS.498.2544H} {498, 2544}

\bibitem[\protect\citeauthoryear{{Javadi}, {van Loon}, {Khosroshahi},
  {Tabatabaei}, {Hamedani Golshan}  \& {Rashidi}}{{Javadi}
  et~al.}{2017}]{Javadi_2017}
{Javadi} A.,  {van Loon} J.~T.,  {Khosroshahi} H.~G.,  {Tabatabaei} F.,
  {Hamedani Golshan} R.,   {Rashidi} M.,  2017, \mn@doi [\mnras]
  {10.1093/mnras/stw2463}, \href
  {https://ui.adsabs.harvard.edu/abs/2017MNRAS.464.2103J} {464, 2103}

\bibitem[\protect\citeauthoryear{{Javanmardi} \& {Kroupa}}{{Javanmardi} \&
  {Kroupa}}{2017}]{Javanmardi_2017}
{Javanmardi} B.,  {Kroupa} P.,  2017, \mn@doi [\aap]
  {10.1051/0004-6361/201629408}, \href
  {https://ui.adsabs.harvard.edu/abs/2017A&A...597A.120J} {597, A120}

\bibitem[\protect\citeauthoryear{{Javanmardi}, {Porciani}, {Kroupa}  \&
  {Pflamm-Altenburg}}{{Javanmardi} et~al.}{2015}]{Javanmardi_2015}
{Javanmardi} B.,  {Porciani} C.,  {Kroupa} P.,   {Pflamm-Altenburg} J.,  2015,
  \mn@doi [\apj] {10.1088/0004-637X/810/1/47}, \href
  {https://ui.adsabs.harvard.edu/abs/2015ApJ...810...47J} {810, 47}

\bibitem[\protect\citeauthoryear{{Jia}, {Hu}  \& {Wang}}{{Jia}
  et~al.}{2023}]{Jia_2023}
{Jia} X.~D.,  {Hu} J.~P.,   {Wang} F.~Y.,  2023, \mn@doi [\aap]
  {10.1051/0004-6361/202346356}, \href
  {https://ui.adsabs.harvard.edu/abs/2023A&A...674A..45J} {674, A45}

\bibitem[\protect\citeauthoryear{{Karachentsev}}{{Karachentsev}}{2012}]{Karachentsev_2012}
{Karachentsev} I.~D.,  2012, \mn@doi [Astrophysical Bulletin]
  {10.1134/S1990341312020010}, \href
  {https://ui.adsabs.harvard.edu/abs/2012AstBu..67..123K} {67, 123}

\bibitem[\protect\citeauthoryear{{Karachentsev} \& {Kaisina}}{{Karachentsev} \&
  {Kaisina}}{2013}]{Karachentsev_2013_SFRproperties}
{Karachentsev} I.~D.,  {Kaisina} E.~I.,  2013, \mn@doi [\aj]
  {10.1088/0004-6256/146/3/46}, \href
  {https://ui.adsabs.harvard.edu/abs/2013AJ....146...46K} {146, 46}

\bibitem[\protect\citeauthoryear{{Karachentsev} \& {Telikova}}{{Karachentsev}
  \& {Telikova}}{2018}]{Karachentsev_2018}
{Karachentsev} I.~D.,  {Telikova} K.~N.,  2018, \mn@doi [Astronomische
  Nachrichten] {10.1002/asna.201813520}, \href
  {https://ui.adsabs.harvard.edu/abs/2018AN....339..615K} {339, 615}

\bibitem[\protect\citeauthoryear{{Karachentsev}, {Karachentseva}, {Huchtmeier}
  \& {Makarov}}{{Karachentsev} et~al.}{2004}]{Karachentsev_2004}
{Karachentsev} I.~D.,  {Karachentseva} V.~E.,  {Huchtmeier} W.~K.,   {Makarov}
  D.~I.,  2004, \mn@doi [\aj] {10.1086/382905}, \href
  {https://ui.adsabs.harvard.edu/abs/2004AJ....127.2031K} {127, 2031}

\bibitem[\protect\citeauthoryear{{Karachentsev}, {Makarov}  \&
  {Kaisina}}{{Karachentsev} et~al.}{2013}]{Karachentsev_2013}
{Karachentsev} I.~D.,  {Makarov} D.~I.,   {Kaisina} E.~I.,  2013, \mn@doi [\aj]
  {10.1088/0004-6256/145/4/101}, \href
  {https://ui.adsabs.harvard.edu/abs/2013AJ....145..101K} {145, 101}

\bibitem[\protect\citeauthoryear{{Keenan}, {Barger}  \& {Cowie}}{{Keenan}
  et~al.}{2013}]{Keenan_2013}
{Keenan} R.~C.,  {Barger} A.~J.,   {Cowie} L.~L.,  2013, \mn@doi [\apj]
  {10.1088/0004-637X/775/1/62}, \href
  {https://ui.adsabs.harvard.edu/abs/2013ApJ...775...62K} {775, 62}

\bibitem[\protect\citeauthoryear{{Kim} et~al.,}{{Kim} et~al.}{2022}]{Kim_2022}
{Kim} J.,  et~al., 2022, \mn@doi [arXiv e-prints]
  {10.48550/arXiv.2212.14539v2}, \href
  {https://ui.adsabs.harvard.edu/abs/2022arXiv221214539K} {p.
  arXiv:2212.14539v2}

\bibitem[\protect\citeauthoryear{{Krishnan}, {Colg{\'a}in}, {Ruchika},
  {Sheikh-Jabbari}  \& {Yang}}{{Krishnan} et~al.}{2020}]{Krishnan_2020}
{Krishnan} C.,  {Colg{\'a}in} E.~{\'O}.,  {Ruchika} Sen A.~A.,
  {Sheikh-Jabbari} M.~M.,   {Yang} T.,  2020, \mn@doi [\prd]
  {10.1103/PhysRevD.102.103525}, \href
  {https://ui.adsabs.harvard.edu/abs/2020PhRvD.102j3525K} {102, 103525}

\bibitem[\protect\citeauthoryear{{Kroupa}}{{Kroupa}}{2001}]{Kroupa_2001}
{Kroupa} P.,  2001, \mn@doi [\mnras] {10.1046/j.1365-8711.2001.04022.x}, \href
  {https://ui.adsabs.harvard.edu/abs/2001MNRAS.322..231K} {322, 231}

\bibitem[\protect\citeauthoryear{{Kroupa} \& {Jerabkova}}{{Kroupa} \&
  {Jerabkova}}{2021}]{Kroupa_Jerabkova_2021}
{Kroupa} P.,  {Jerabkova} T.,  2021, arXiv e-prints, \href
  {https://ui.adsabs.harvard.edu/abs/2021arXiv211210788K} {p.
  arXiv:2112.10788v1}

\bibitem[\protect\citeauthoryear{{Kroupa} \& {Weidner}}{{Kroupa} \&
  {Weidner}}{2003}]{Kroupa_2003}
{Kroupa} P.,  {Weidner} C.,  2003, \mn@doi [\apj] {10.1086/379105}, \href
  {https://ui.adsabs.harvard.edu/abs/2003ApJ...598.1076K} {598, 1076}

\bibitem[\protect\citeauthoryear{{Kroupa}, {Haslbauer}, {Banik}, {Nagesh}  \&
  {Pflamm-Altenburg}}{{Kroupa} et~al.}{2020}]{Kroupa_2020}
{Kroupa} P.,  {Haslbauer} M.,  {Banik} I.,  {Nagesh} S.~T.,
  {Pflamm-Altenburg} J.,  2020, \mn@doi [\mnras] {10.1093/mnras/staa1851},
  \href {https://ui.adsabs.harvard.edu/abs/2020MNRAS.497...37K} {497, 37}

\bibitem[\protect\citeauthoryear{{Laporte}, {Meyer}, {Ellis}, {Robertson},
  {Chisholm}  \& {Roberts-Borsani}}{{Laporte} et~al.}{2021}]{Laporte_2021}
{Laporte} N.,  {Meyer} R.~A.,  {Ellis} R.~S.,  {Robertson} B.~E.,  {Chisholm}
  J.,   {Roberts-Borsani} G.~W.,  2021, \mn@doi [\mnras]
  {10.1093/mnras/stab1239}, \href
  {https://ui.adsabs.harvard.edu/abs/2021MNRAS.505.3336L} {505, 3336}

\bibitem[\protect\citeauthoryear{{Leja}, {van Dokkum}, {Franx}  \&
  {Whitaker}}{{Leja} et~al.}{2015}]{Leja_2015}
{Leja} J.,  {van Dokkum} P.~G.,  {Franx} M.,   {Whitaker} K.~E.,  2015, \mn@doi
  [\apj] {10.1088/0004-637X/798/2/115}, \href
  {https://ui.adsabs.harvard.edu/abs/2015ApJ...798..115L} {798, 115}

\bibitem[\protect\citeauthoryear{{Leja}, {Carnall}, {Johnson}, {Conroy}  \&
  {Speagle}}{{Leja} et~al.}{2019}]{Leja_2019}
{Leja} J.,  {Carnall} A.~C.,  {Johnson} B.~D.,  {Conroy} C.,   {Speagle} J.~S.,
   2019, \mn@doi [\apj] {10.3847/1538-4357/ab133c}, \href
  {https://ui.adsabs.harvard.edu/abs/2019ApJ...876....3L} {876, 3}

\bibitem[\protect\citeauthoryear{{Lilly}, {Le Fevre}, {Hammer}  \&
  {Crampton}}{{Lilly} et~al.}{1996}]{Lilly_1996}
{Lilly} S.~J.,  {Le Fevre} O.,  {Hammer} F.,   {Crampton} D.,  1996, \mn@doi
  [\apjl] {10.1086/309975}, \href
  {https://ui.adsabs.harvard.edu/abs/1996ApJ...460L...1L} {460, L1}

\bibitem[\protect\citeauthoryear{{Madau} \& {Dickinson}}{{Madau} \&
  {Dickinson}}{2014}]{Madau_Dickinson_2014}
{Madau} P.,  {Dickinson} M.,  2014, \mn@doi [\araa]
  {10.1146/annurev-astro-081811-125615}, \href
  {https://ui.adsabs.harvard.edu/abs/2014ARA&A..52..415M} {52, 415}

\bibitem[\protect\citeauthoryear{{Madau} \& {Fragos}}{{Madau} \&
  {Fragos}}{2017}]{Madau_2017}
{Madau} P.,  {Fragos} T.,  2017, \mn@doi [\apj] {10.3847/1538-4357/aa6af9},
  \href {https://ui.adsabs.harvard.edu/abs/2017ApJ...840...39M} {840, 39}

\bibitem[\protect\citeauthoryear{{Madau}, {Ferguson}, {Dickinson},
  {Giavalisco}, {Steidel}  \& {Fruchter}}{{Madau} et~al.}{1996}]{Madau_1996}
{Madau} P.,  {Ferguson} H.~C.,  {Dickinson} M.~E.,  {Giavalisco} M.,  {Steidel}
  C.~C.,   {Fruchter} A.,  1996, \mn@doi [\mnras] {10.1093/mnras/283.4.1388},
  \href {https://ui.adsabs.harvard.edu/abs/1996MNRAS.283.1388M} {283, 1388}

\bibitem[\protect\citeauthoryear{{Madau}, {Pozzetti}  \& {Dickinson}}{{Madau}
  et~al.}{1998}]{Madau_1998}
{Madau} P.,  {Pozzetti} L.,   {Dickinson} M.,  1998, \mn@doi [\apj]
  {10.1086/305523}, \href
  {https://ui.adsabs.harvard.edu/abs/1998ApJ...498..106M} {498, 106}

\bibitem[\protect\citeauthoryear{{Maddox}, {Sutherland}, {Efstathiou},
  {Loveday}  \& {Peterson}}{{Maddox} et~al.}{1990}]{Maddox_1990}
{Maddox} S.~J.,  {Sutherland} W.~J.,  {Efstathiou} G.,  {Loveday} J.,
  {Peterson} B.~A.,  1990, \mnras, \href
  {https://ui.adsabs.harvard.edu/abs/1990MNRAS.247P...1M} {247, 1P}

\bibitem[\protect\citeauthoryear{{Massana} et~al.,}{{Massana}
  et~al.}{2022}]{Massana_2022}
{Massana} P.,  et~al., 2022, \mn@doi [\mnras] {10.1093/mnrasl/slac030}, \href
  {https://ui.adsabs.harvard.edu/abs/2022MNRAS.513L..40M} {513, L40}

\bibitem[\protect\citeauthoryear{{McGaugh} \& {Schombert}}{{McGaugh} \&
  {Schombert}}{2014}]{McGaugh_2014}
{McGaugh} S.~S.,  {Schombert} J.~M.,  2014, \mn@doi [\aj]
  {10.1088/0004-6256/148/5/77}, \href
  {https://ui.adsabs.harvard.edu/abs/2014AJ....148...77M} {148, 77}

\bibitem[\protect\citeauthoryear{{McGaugh}, {Schombert}  \& {Lelli}}{{McGaugh}
  et~al.}{2017}]{McGaugh_2017}
{McGaugh} S.~S.,  {Schombert} J.~M.,   {Lelli} F.,  2017, \mn@doi [\apj]
  {10.3847/1538-4357/aa9790}, \href
  {https://ui.adsabs.harvard.edu/abs/2017ApJ...851...22M} {851, 22}

\bibitem[\protect\citeauthoryear{{Migkas}, {Schellenberger}, {Reiprich},
  {Pacaud}, {Ramos-Ceja}  \& {Lovisari}}{{Migkas} et~al.}{2020}]{Migkas_2020}
{Migkas} K.,  {Schellenberger} G.,  {Reiprich} T.~H.,  {Pacaud} F.,
  {Ramos-Ceja} M.~E.,   {Lovisari} L.,  2020, \mn@doi [\aap]
  {10.1051/0004-6361/201936602}, \href
  {https://ui.adsabs.harvard.edu/abs/2020A&A...636A..15M} {636, A15}

\bibitem[\protect\citeauthoryear{{Migkas}, {Pacaud}, {Schellenberger}, {Erler},
  {Nguyen-Dang}, {Reiprich}, {Ramos-Ceja}  \& {Lovisari}}{{Migkas}
  et~al.}{2021}]{Migkas_2021}
{Migkas} K.,  {Pacaud} F.,  {Schellenberger} G.,  {Erler} J.,  {Nguyen-Dang}
  N.~T.,  {Reiprich} T.~H.,  {Ramos-Ceja} M.~E.,   {Lovisari} L.,  2021,
  \mn@doi [\aap] {10.1051/0004-6361/202140296}, \href
  {https://ui.adsabs.harvard.edu/abs/2021A&A...649A.151M} {649, A151}

\bibitem[\protect\citeauthoryear{{Milgrom}}{{Milgrom}}{1983}]{Milgrom_1983}
{Milgrom} M.,  1983, \mn@doi [\apj] {10.1086/161130}, \href
  {https://ui.adsabs.harvard.edu/abs/1983ApJ...270..365M} {270, 365}

\bibitem[\protect\citeauthoryear{{Peebles} \& {Nusser}}{{Peebles} \&
  {Nusser}}{2010}]{Peebles_2010}
{Peebles} P.~J.~E.,  {Nusser} A.,  2010, \mn@doi [\nat] {10.1038/nature09101},
  \href {https://ui.adsabs.harvard.edu/abs/2010Natur.465..565P} {465, 565}

\bibitem[\protect\citeauthoryear{{Pipino}, {Lilly}  \& {Carollo}}{{Pipino}
  et~al.}{2014}]{Pipino_2014}
{Pipino} A.,  {Lilly} S.~J.,   {Carollo} C.~M.,  2014, \mn@doi [\mnras]
  {10.1093/mnras/stu579}, \href
  {https://ui.adsabs.harvard.edu/abs/2014MNRAS.441.1444P} {441, 1444}

\bibitem[\protect\citeauthoryear{{Planck Collaboration XIII}}{{Planck
  Collaboration XIII}}{2016}]{Planck_2016_IllustrisTNG}
{Planck Collaboration XIII} 2016, \mn@doi [\aap] {10.1051/0004-6361/201525830},
  \href {https://ui.adsabs.harvard.edu/abs/2016A\&A...594A..13P} {594, A13}

\bibitem[\protect\citeauthoryear{{Reddy}, {Pettini}, {Steidel}, {Shapley},
  {Erb}  \& {Law}}{{Reddy} et~al.}{2012}]{Reddy_2012}
{Reddy} N.~A.,  {Pettini} M.,  {Steidel} C.~C.,  {Shapley} A.~E.,  {Erb} D.~K.,
    {Law} D.~R.,  2012, \mn@doi [\apj] {10.1088/0004-637X/754/1/25}, \href
  {https://ui.adsabs.harvard.edu/abs/2012ApJ...754...25R} {754, 25}

\bibitem[\protect\citeauthoryear{{Riechers} et~al.,}{{Riechers}
  et~al.}{2019}]{Riechers_2019}
{Riechers} D.~A.,  et~al., 2019, \mn@doi [\apj] {10.3847/1538-4357/aafc27},
  \href {https://ui.adsabs.harvard.edu/abs/2019ApJ...872....7R} {872, 7}

\bibitem[\protect\citeauthoryear{{Rubart} \& {Schwarz}}{{Rubart} \&
  {Schwarz}}{2013}]{Rubart_Schwarz_2013}
{Rubart} M.,  {Schwarz} D.~J.,  2013, \mn@doi [\aap]
  {10.1051/0004-6361/201321215}, \href
  {https://ui.adsabs.harvard.edu/abs/2013A&A...555A.117R} {555, A117}

\bibitem[\protect\citeauthoryear{{Rubart}, {Bacon}  \& {Schwarz}}{{Rubart}
  et~al.}{2014}]{Rubart_2014}
{Rubart} M.,  {Bacon} D.,   {Schwarz} D.~J.,  2014, \mn@doi [\aap]
  {10.1051/0004-6361/201423583}, \href
  {https://ui.adsabs.harvard.edu/abs/2014A&A...565A.111R} {565, A111}

\bibitem[\protect\citeauthoryear{{Salpeter}}{{Salpeter}}{1955}]{Salpeter_1955}
{Salpeter} E.~E.,  1955, \mn@doi [\apj] {10.1086/145971}, \href
  {https://ui.adsabs.harvard.edu/abs/1955ApJ...121..161S} {121, 161}

\bibitem[\protect\citeauthoryear{{Sanders}}{{Sanders}}{1998}]{Sanders_1998}
{Sanders} R.~H.,  1998, \mn@doi [\mnras] {10.1046/j.1365-8711.1998.01459.x},
  \href {https://ui.adsabs.harvard.edu/abs/1998MNRAS.296.1009S} {296, 1009}

\bibitem[\protect\citeauthoryear{{Schombert}, {McGaugh}  \&
  {Lelli}}{{Schombert} et~al.}{2019}]{Schombert_2019}
{Schombert} J.,  {McGaugh} S.,   {Lelli} F.,  2019, \mn@doi [\mnras]
  {10.1093/mnras/sty3223}, \href
  {https://ui.adsabs.harvard.edu/abs/2019MNRAS.483.1496S} {483, 1496}

\bibitem[\protect\citeauthoryear{{Schwarz}, {Copi}, {Huterer}  \&
  {Starkman}}{{Schwarz} et~al.}{2016}]{Schwarz_2016}
{Schwarz} D.~J.,  {Copi} C.~J.,  {Huterer} D.,   {Starkman} G.~D.,  2016,
  \mn@doi [Classical and Quantum Gravity] {10.1088/0264-9381/33/18/184001},
  \href {https://ui.adsabs.harvard.edu/abs/2016CQGra..33r4001S} {33, 184001}

\bibitem[\protect\citeauthoryear{{Secrest}, {von Hausegger}, {Rameez},
  {Mohayaee}, {Sarkar}  \& {Colin}}{{Secrest} et~al.}{2021}]{Secrest_2021}
{Secrest} N.~J.,  {von Hausegger} S.,  {Rameez} M.,  {Mohayaee} R.,  {Sarkar}
  S.,   {Colin} J.,  2021, \mn@doi [\apjl] {10.3847/2041-8213/abdd40}, \href
  {https://ui.adsabs.harvard.edu/abs/2021ApJ...908L..51S} {908, L51}

\bibitem[\protect\citeauthoryear{{Secrest}, {von Hausegger}, {Rameez},
  {Mohayaee}  \& {Sarkar}}{{Secrest} et~al.}{2022}]{Secrest_2022}
{Secrest} N.~J.,  {von Hausegger} S.,  {Rameez} M.,  {Mohayaee} R.,   {Sarkar}
  S.,  2022, \mn@doi [\apjl] {10.3847/2041-8213/ac88c0}, \href
  {https://ui.adsabs.harvard.edu/abs/2022ApJ...937L..31S} {937, L31}

\bibitem[\protect\citeauthoryear{{Speagle}, {Steinhardt}, {Capak}  \&
  {Silverman}}{{Speagle} et~al.}{2014}]{Speagle_2014}
{Speagle} J.~S.,  {Steinhardt} C.~L.,  {Capak} P.~L.,   {Silverman} J.~D.,
  2014, \mn@doi [\apjs] {10.1088/0067-0049/214/2/15}, \href
  {https://ui.adsabs.harvard.edu/abs/2014ApJS..214...15S} {214, 15}

\bibitem[\protect\citeauthoryear{{Tomczak} et~al.,}{{Tomczak}
  et~al.}{2016}]{Tomczak_2016}
{Tomczak} A.~R.,  et~al., 2016, \mn@doi [\apj] {10.3847/0004-637X/817/2/118},
  \href {https://ui.adsabs.harvard.edu/abs/2016ApJ...817..118T} {817, 118}

\bibitem[\protect\citeauthoryear{{Walter} et~al.,}{{Walter}
  et~al.}{2020}]{Walter_2020}
{Walter} F.,  et~al., 2020, \mn@doi [\apj] {10.3847/1538-4357/abb82e}, \href
  {https://ui.adsabs.harvard.edu/abs/2020ApJ...902..111W} {902, 111}

\bibitem[\protect\citeauthoryear{{Weisz}, {Dolphin}, {Skillman}, {Holtzman},
  {Dalcanton}, {Cole}  \& {Neary}}{{Weisz} et~al.}{2013}]{Weisz_2013}
{Weisz} D.~R.,  {Dolphin} A.~E.,  {Skillman} E.~D.,  {Holtzman} J.,
  {Dalcanton} J.~J.,  {Cole} A.~A.,   {Neary} K.,  2013, \mn@doi [\mnras]
  {10.1093/mnras/stt165}, \href
  {https://ui.adsabs.harvard.edu/abs/2013MNRAS.431..364W} {431, 364}

\bibitem[\protect\citeauthoryear{{Weisz}, {Dolphin}, {Skillman}, {Holtzman},
  {Gilbert}, {Dalcanton}  \& {Williams}}{{Weisz} et~al.}{2014}]{Weisz_2014}
{Weisz} D.~R.,  {Dolphin} A.~E.,  {Skillman} E.~D.,  {Holtzman} J.,  {Gilbert}
  K.~M.,  {Dalcanton} J.~J.,   {Williams} B.~F.,  2014, \mn@doi [\apj]
  {10.1088/0004-637X/789/2/148}, \href
  {https://ui.adsabs.harvard.edu/abs/2014ApJ...789..148W} {789, 148}

\bibitem[\protect\citeauthoryear{{Wilkins}, {Trentham}  \& {Hopkins}}{{Wilkins}
  et~al.}{2008}]{Wilkins_2008}
{Wilkins} S.~M.,  {Trentham} N.,   {Hopkins} A.~M.,  2008, \mn@doi [\mnras]
  {10.1111/j.1365-2966.2008.12885.x}, \href
  {https://ui.adsabs.harvard.edu/abs/2008MNRAS.385..687W} {385, 687}

\bibitem[\protect\citeauthoryear{{Wiltshire}}{{Wiltshire}}{2007a}]{Wiltshire_2007a}
{Wiltshire} D.~L.,  2007a, \mn@doi [New Journal of Physics]
  {10.1088/1367-2630/9/10/377}, \href
  {https://ui.adsabs.harvard.edu/abs/2007NJPh....9..377W} {9, 377}

\bibitem[\protect\citeauthoryear{{Wiltshire}}{{Wiltshire}}{2007b}]{Wiltshire_2007b}
{Wiltshire} D.~L.,  2007b, \mn@doi [\prl] {10.1103/PhysRevLett.99.251101},
  \href {https://ui.adsabs.harvard.edu/abs/2007PhRvL..99y1101W} {99, 251101}

\bibitem[\protect\citeauthoryear{{Wiltshire}}{{Wiltshire}}{2009}]{Wiltshire_2009}
{Wiltshire} D.~L.,  2009, \mn@doi [\prd] {10.1103/PhysRevD.80.123512}, \href
  {https://ui.adsabs.harvard.edu/abs/2009PhRvD..80l3512W} {80, 123512}

\bibitem[\protect\citeauthoryear{{Wiltshire}}{{Wiltshire}}{2019}]{Wiltshire_2019}
{Wiltshire} D.~L.,  2019, \mn@doi [\aap] {10.1051/0004-6361/201834833}, \href
  {https://ui.adsabs.harvard.edu/abs/2019A&A...624A..12W} {624, A12}

\bibitem[\protect\citeauthoryear{{Wong} et~al.,}{{Wong}
  et~al.}{2020}]{Wong_2020}
{Wong} K.~C.,  et~al., 2020, \mn@doi [\mnras] {10.1093/mnras/stz3094}, \href
  {https://ui.adsabs.harvard.edu/abs/2020MNRAS.498.1420W} {498, 1420}

\bibitem[\protect\citeauthoryear{{Wong}, {Shanks}, {Metcalfe}  \&
  {Whitbourn}}{{Wong} et~al.}{2022}]{Wong_2022}
{Wong} J. H.~W.,  {Shanks} T.,  {Metcalfe} N.,   {Whitbourn} J.~R.,  2022,
  \mn@doi [\mnras] {10.1093/mnras/stac396}, \href
  {https://ui.adsabs.harvard.edu/abs/2022MNRAS.511.5742W} {511, 5742}

\bibitem[\protect\citeauthoryear{{Yadav}, {Bagla}  \& {Khandai}}{{Yadav}
  et~al.}{2010}]{Yadav_2010}
{Yadav} J.~K.,  {Bagla} J.~S.,   {Khandai} N.,  2010, \mn@doi [\mnras]
  {10.1111/j.1365-2966.2010.16612.x}, \href
  {https://ui.adsabs.harvard.edu/abs/2010MNRAS.405.2009Y} {405, 2009}

\bibitem[\protect\citeauthoryear{{Yan}, {Jerabkova}  \& {Kroupa}}{{Yan}
  et~al.}{2019}]{Yan_2019b}
{Yan} Z.,  {Jerabkova} T.,   {Kroupa} P.,  2019, \mn@doi [\aap]
  {10.1051/0004-6361/201936636}, \href
  {https://ui.adsabs.harvard.edu/abs/2019A&A...632A.110Y} {632, A110}

\bibitem[\protect\citeauthoryear{{Yu} \& {Wang}}{{Yu} \&
  {Wang}}{2016}]{Yu_2016}
{Yu} H.,  {Wang} F.~Y.,  2016, \mn@doi [\apj] {10.3847/0004-637X/820/2/114},
  \href {https://ui.adsabs.harvard.edu/abs/2016ApJ...820..114Y} {820, 114}

\bibitem[\protect\citeauthoryear{{Zucca} et~al.,}{{Zucca}
  et~al.}{1997}]{Zucca_1997}
{Zucca} E.,  et~al., 1997, \mn@doi [\aap] {10.48550/arXiv.astro-ph/9705096},
  \href {https://ui.adsabs.harvard.edu/abs/1997A&A...326..477Z} {326, 477}

\makeatother
\end{thebibliography}



\bsp
\label{lastpage} 
\end{document}